\documentstyle[psfig]{mn} 
\input mnextra.sty
\def\vp{{\caps vp}}

\def\vpar{v_{\scriptscriptstyle \Vert}}

\def\encapfig#1#2#3#4{\hspace*{#4}
        {\psfig{figure=#1,width=#2,angle=#3} } }

\newif\ifpsfiles\psfilestrue
\def\getfig#1{\ifpsfiles\psfig{figure=#1,width=\hsize}\fi}

\def\hfifty{h_{50}}

\title[The dark halo of the E0 galaxy NGC 6703]
	{Breaking the degeneracy between anisotropy and mass:
	The dark halo of the E0 galaxy NGC 6703}

\author[O.E.\ Gerhard, G.\ Jeske, R.P.\ Saglia, R.\ Bender] 
       {Ortwin Gerhard$^{1,2}$, Gunther Jeske$^2$, 
        R.\ P.\ Saglia$^{3,}$\thanks{Visiting Astronomer of 
	the German-Spanish Astronomical Center, Calar Alto, 
	operated by the Max Plank Institut f\"ur Astronomie, 
	Heidelberg, jointly with the Spanish National Commission 
	for Astronomy}, Ralf Bender$^{3,\mbox{$\star$}}$  \\ 
$^1$ Astronomisches Institut, Universit\"at Basel, Venusstrasse 7,
     CH-4102 Binningen, Switzerland \\
$^2$ Landessternwarte, K\"onigstuhl, D-69117 Heidelberg, Germany \\ 
$^3$ Institut f\"ur Astronomie und Astrophysik, Scheinerstr. 1, D-81679 
     M\"unchen, Germany
}
\date{
Accepted October 1997. Received May 1997, 
in original form June 1995.
} 
\begin{document} 
\maketitle
\begin{abstract} 

We have measured line-of-sight velocity profiles (\vp s) in the E0
galaxy NGC 6703 out to $2.6 R_e$. Comparing with the \vp s predicted
from spherical distribution functions (\df s), we constrain the mass
distribution and the anisotropy of the stellar orbits in this galaxy.

\hspace{5mm}
We have developed a non-parametric technique to determine the \df\
$f(E,L^2)$ directly from the kinematic data. We test this technique on
Monte Carlo simulated data with the spatial extent, sampling, and
error bars of the NGC 6703 data.  We find that smooth underlying \df s
can be recovered to an \rms\ accuracy of $12\%$ inside three times the
radius of the last kinematic data point, and the anisotropy parameter
$\beta(r)$ to an accuracy of $0.1$, in a {\sl known} potential. These
uncertainties can be reduced with improved data.

\hspace{5mm}
By comparing such best-estimate, regularized models in different
potentials, we can derive constraints on the mass distribution and
anisotropy. Tests show that with presently available data, an
asymptotically constant halo circular velocity $v_0$ can be determined
with an accuracy of $\pm \lta 50\kms$. This formal range often
includes high--$v_0$ models with implausibly large gradients across
the data boundary. However, even with extremely high quality data some
uncertainty on the detailed shape of the underlying circular velocity
curve remains.

\hspace{5mm}
In the case of NGC 6703 we thus determine the true circular velocity
at $2.6 R_e$ to be $250\pm 40\kms$ at $95\%$ confidence, corresponding
to a total mass in NGC 6703 inside $78''$ ($13.5\,\hfifty^{-1} \kpc$,
where $h_{50}\equiv H_0 / 50 {\rm km/s/Mpc}$) of $1.6-2.6
\times 10^{11} \hfifty^{-1} \msun$.  No model without dark matter will
fit the data; however, a {\sl maximum stellar mass} model in which the
luminous component provides nearly all the mass in the centre does. In
such a model, the total luminous mass inside $78''$ is $9 \times
10^{10} \msun$ and the integrated B-band mass--to--light ratio out to
this radius is $\Upsilon_B=5.3-10$, corresponding to a rise from the
center by at least a factor of $1.6$.

\hspace{5mm}
The anisotropy of the stellar distribution function in NGC 6703
changes from near-isotropic at the centre to slightly radially
anisotropic ($\beta\equal 0.3-0.4$ at 30'', $\beta\equal 0.2-0.4$ at
60'') and is not well-constrained at the outer edge of the data, where
$\beta\equal -0.5 - +0.4$, depending on variations of the potential in
the allowed range.

\hspace{5mm}
Our results suggest that also elliptical galaxies begin to be
dominated by dark matter at radii of $\sim 10\kpc$.

\end{abstract}

\begin{keywords}
stellar dynamics -- dark matter -- galaxies: elliptical and lenticular 
-- galaxies: kinematics and dynamics -- galaxies: individual -- line profiles
\end{keywords}

\section{Introduction}

Current cosmological models predict that, similar to spiral galaxies,
elliptical galaxies should be surrounded by dark matter haloes. The
observational evidence for dark matter in ellipticals is still weak,
however. In a few cases masses have been determined from X-ray
observations (e.g., Awaki \etal 1994, Kim \& Fabbiano 1995) or HI ring
velocities (Franx, van Gorkom \& de Zeeuw 1994).  In others it has
been possible to rule out constant M/L from extended velocity
dispersion data (Saglia \etal 1993), from absorption line profile
measurements (Carollo \etal 1995, Rix et al. 1997), or from globular
cluster or planetary nebula velocities (e.g., Grillmair \etal 1994,
Arnaboldi \etal 1994).  Gravitational lensing statistics (Maoz \& Rix
1993) and individual image lens separations (Kochanek \& Keeton 1997)
favour models with extended dark matter haloes around
ellipticals. Despite of this, the detailed radial mass distribution in
elliptical galaxies remains largely unknown. Similarly, although we
know from the tensor virial theorem that giant ellipticals are
globally anisotropic (Binney 1978), their detailed anisotropy
structure is only poorly known.

The origin of this uncertainty is a fundamental degeneracy -- in
general, it is impossible to disentangle the anisotropy in the
velocity distribution and the gravitational potential from velocity
dispersion and rotation measurements alone (Binney \& Mamon 1982,
Dejonghe \& Merritt 1992).  Tangential anisotropy, for example, can
mimic the presence of dark matter. Recent dynamical studies have
indicated, however, that the anisotropy of the stellar distribution
function (\df) is reflected in the shapes of the line-of-sight
velocity profiles (\vp s) in a way that depends on the gravitational
potential (Gerhard 1993, G93; Merritt 1993). These papers argued that
the extra constraints derived from the \vp\ measurements may be enough
to break the degeneracy and determine the mass distribution.

If this is correct, it provides a new method to investigate the
properties of the dark matter haloes around elliptical galaxies at
intermediate radii: \vp s can now be estimated from high-quality
absorption line measurements out to $\sim 3$ effective radii.
Dynamical models are then used to disentangle the effects of orbital
anisotropy and potential gradient on the \vp\ shapes.  In this paper,
we implement these ideas for the analysis of real data, analyzing the
E0 galaxy NGC 6703.  This study is part of an observational and
theoretical program aimed at understanding the mass distribution and
orbital structure in elliptical galaxies. Preliminary accounts of this
work have been given in Jeske \etal (1996) and Saglia \etal (1997a).

We have obtained long--slit spectroscopy for NGC 6703, and have
measured \vp s to $\sim 2.6 R_e$ with the method of Bender (1990).
The results are quantified by a Gauss-Hermite decomposition (G93, van
der Marel \& Franx 1993) as described by Bender, Saglia \& Gerhard
(1994; BSG). These observations are described in Section 2.  In
Section 3 we use simple dynamical models to describe the variation of
the \vp\ shapes with anisotropy and potential, generalizing the
results of G93 for scale--free models. These models, taken from a
systematic study of the relation between \df\ and \vp s in spherical
potentials (Jeske 1995), are described in Appendix A.  In Section 4 we
develop a non-parametric method for inferring the \df\ and potential
from absorption line profile measurements. Tests on Monte Carlo
generated data are used to determine the degree of confidence with
which the \df\ and potential can be inferred from real data.
In Section 5 we analyse the kinematic data for NGC 6703. Comparison
with the dynamical models from Jeske (1995) already shows that no
constant--M/L model will fit the data. The non--parametric method
developed in Section 4 is then used to derive quantitative constraints
on the mass distribution and anisotropy of this galaxy. Finally,
Section 6 presents a discussion of the results and our conclusions.

\section{Observations of NGC 6703}

NGC 6703 is an E0 galaxy at a distance $D\equal 36$ Mpc (Faber et al.\
1989) for $H_0\equal 50$ km/s/Mpc.  From a B-band CCD frame taken at
the prime focus of the 3.5m telescope on Calar Alto and kindly
provided by U.~Hopp we have measured the inner surface brightness
profile using the algorithm by Saglia et al. (1997b); 
this follows an $R^{1/4}$ law with $R_e \equal 30''\equal
5.2\,\hfifty^{-1}$ kpc, or a Jaffe model with $r_J\equal 46.5''\equal
8.1\,\hfifty^{-1}\kpc$, with small residuals (Fig.~1). A 3\% increase
of the sky value reduces the measured Jaffe radius to $35.5''$, a 1\%
decrease of the sky values increases it to $54''$.
Isophote shapes deviate little from circles ($\epsilon\lt 0.05$, 
$\mid a_4/a\mid<0.005$) and
show small twisting ($\Delta PA \ssimeq 10^\circ$). From the Jaffe
profile fit we derive a fiducial (calibrated and corrected for galactic
absorption following Faber et al. 1989) $M_B\equal -21.07$, or luminosity
$L_B \equal 4.16\times 10^{10} \hfifty^{-2} L_{\odot,B}$.  
Note that the values of $R_e$ and $M_B$ derived here are slightly larger than
those ($R_e=24''$, $M_B=-20.79$) given by Faber et al. (1989).

The spectroscopic observations were carried out in October 1994, May
1995 and August 1995 with the 3.5-m telescope on Calar Alto, Spain. In
all of the runs the same setup was adopted. The Boller \& Chivens
longslit twin spectrograph was used with a 1200 line mm$^{-1}$ grating
giving 36 \AA/mm dispersion. The detector was a Tektronix CCD with
1024$\times$1024 24$\mu$m pixels and the wavelength range 4760-5640
\AA.  The instrumental resolution obtained using a 3.6 arcsec wide
slit was 85 km/s. We collected 1.5 hrs of observations along the major
axis of the galaxy, 4 hrs of observations perpendicular to the major
axis and shifted to the North-East of the center by 24 arcsec ($0.8
R_e$), and 13 hours of observations perpendicular to the major axis
and shifted to the North-East of the center by 36 arcsec ($1.2 R_e$).
Spectra taken parallel to the minor axis and shifted from the center
allow at the same time a good sky subtraction and the symmetry check of 
the data points.

\begin{figure}
\vspace*{-0.35cm}
\getfig{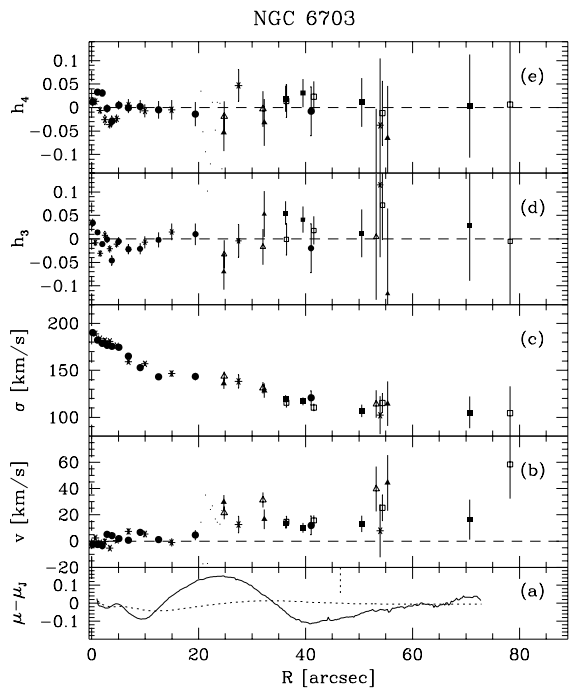}
 \vspace*{-0.8cm}
 \caption[Photometry and kinematics of NGC 6703]
    {(a) Residuals of a Jaffe model fit to photometry for NGC 6703. Full
      line: in surface brightness; dotted line: in the curve of growth.
      The vertical dotted line marks the scaling Jaffe radius $r_J$.
      (b) Folded mean velocity, (c) velocity dispersion, (d) $h_3$, and (e)
      $h_4$ profiles. Crosses and filled circles refer to the two
      sides of the galaxy and the major axis spectrum. The small dots refer
      to the unrebinned spectrum (see text). Open and filled 
      triangles refer to the two sides of the galaxy and the spectrum taken 
      parallel to the minor axis and shifted 24 arcsec from the center.
      Open and filled squares refer to the two sides of the galaxy and the
      spectrum taken parallel to the minor axis and shifted 36 arcsec from 
      the center. }
 \label{obsvns}
\end{figure}

The analysis of the data was carried out following the steps described
by BSG. The logarithmic wavelength calibration was performed at a
smaller step ($\Delta v=30$ km/s) than the actual pixel size ($\approx
50$ km/s) to exploit the full capabilities of the Fourier Correlation
Quotient method.  A sky subtraction better than 1\% was achieved.  The
heliocentric velocity difference between the May 1995 and the
September 1994 - August 1995 frames was taken into account before
coadding the observed spectra. The spectra were rebinned along the
spatial direction to obtain a nearly constant signal to noise ratio
larger than 50 per resolution element.  The effects of the continuum
fitting and instrumental resolution were extensively tested by Monte
Carlo simulations.  The residual systematic effects on the values of
the $h_3$ and $h_4$ parameters are expected to be less than 0.01. The
resulting fitted values for the folded velocity $v$ , velocity
dispersion $\sigma$, $h_3$ and $h_4$ profiles are shown in
Fig.~\ref{obsvns} as a function of the distance from the center,
reaching $\sim 2.6 R_e$. The $v$ and $\sigma$ profiles are folded
antisymetrically with respect to the center for the major axis
spectrum, symmetrically with respect to the major axis for the spectra
parallel to the minor axis. Template mismatching was minimized by
choosing the template star which gave the minimal symmetric $h_3$
profile derived along the major axis of the galaxy.  The systematic
effect due to the residual mismatching on the derived $h_4$ values
estimated from the remaining symmetric components is less than 0.01.

The galaxy shows very little rotation ($\approx 0$ km/s for $R<R_e$,
$\approx 20-30$ km/s for $R>R_e$). The (cylindrical) rotation measured
parallel to the minor axis is slightly larger ($\approx 35$ km/s)
along the 24 arcsec shifted spectrum, but consistent with the peak
velocity reached along the major axis at $R\approx 22$ arcsec. This is
shown by the velocities derived from the unbinned major axis spectra
(dots in Fig.~\ref{obsvns}). The velocity dispersion drops from the central
$\approx\!  190$ km/s to $\approx\!  140$ km/s at $R_e/2$, slowly
declining to about 110 km/s in the outer parts. The $h_3$ and $h_4$
values are everywhere close to zero.  The error bars are determined
from Monte-Carlo simulations. Noise is added to  template stars 
(rebinned to the original wavelength pixel size) broadened following the
observed values of $\sigma$ and $h_4$, matching the power spectrum
noise to peak ratio of the galaxy spectra. The accuracy of the estimated 
error bars (the r.m.s. of 30 replica of the data points) is about 20 \%
(determined from the scatter of the estimated signal to noise ratios).

Data points at the largest distances for the different data sets have
lower signal to noise ratio than the mean and therefore have larger
error bars.  In addition, these data points are expected to suffer
more from the systematic effects due to the galaxy light contamination
of the sky subtraction(see discussion in Saglia et al. 1993). They are
in any case consistent within the error bars with the more accurate
values derived from the other available spectra.

The observed  scatter is sometimes slightly larger than expected from
the error estimates. This excess could be real and due to the faint
structures apparent in an unsharped mask image of the galaxy. In
particular this applies to the asymmetries observed in the $h_4$
profile in the central 5 arcsec. The negative $h_4$ values detected
for the first data points of the 24 arcsec shifted spectrum are also
real. They are detected in the unbinned major axis spectra at
$R\approx 22$ arcsec (dots in Fig.~\ref{obsvns}).


\begin{figure*}
\vspace*{0cm}
\centering
\encapfig{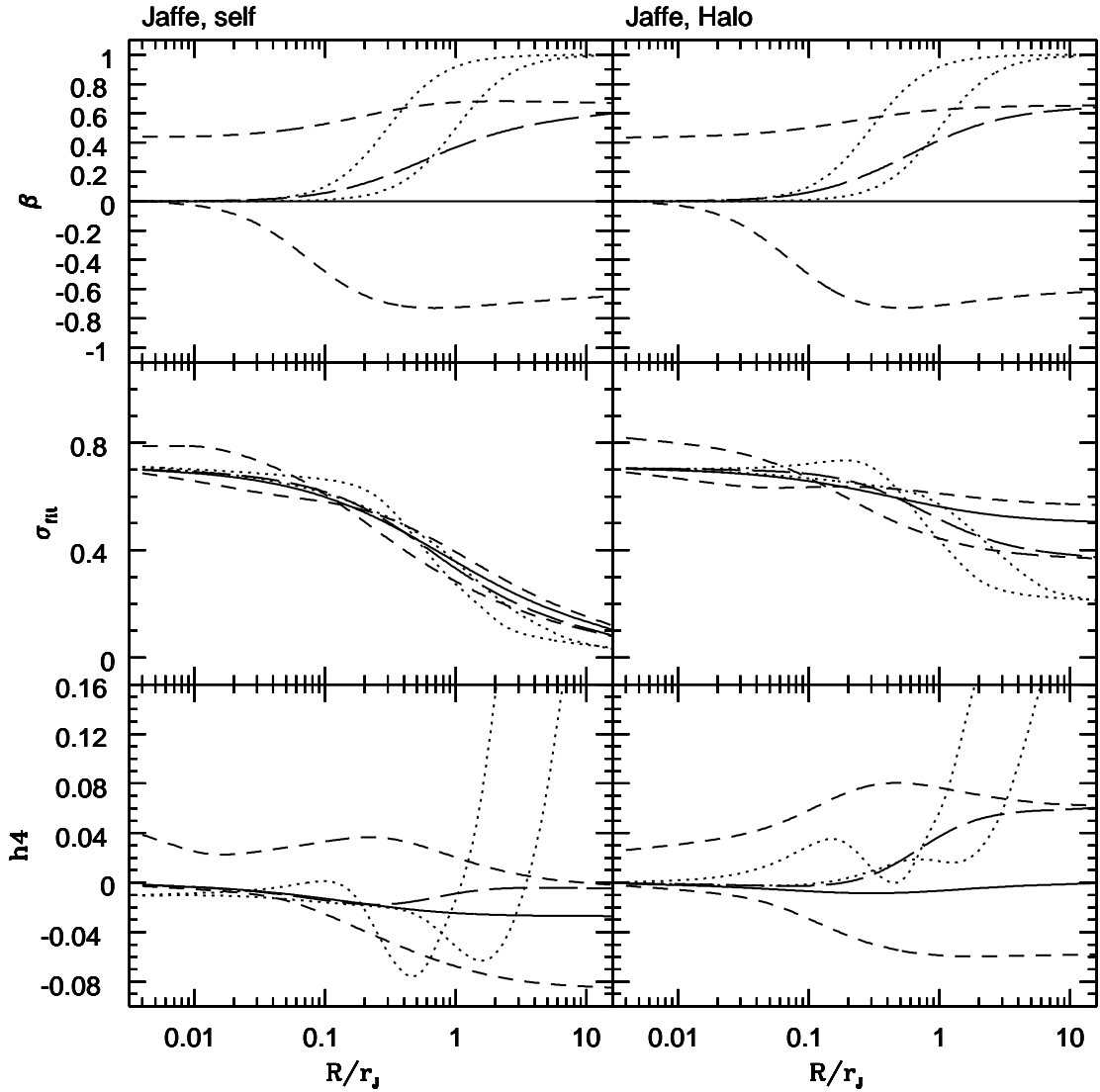}{15.5cm}{0}{0cm}
 \vspace*{-0.4cm}
 \caption[models]
    {Fitted projected velocity dispersion $\sigma_{\rm fit}$,
     anisotropy parameter $\beta$,
     and \vp-parameter $h_4$ for
     representative Jaffe models in self-consistent (left) and
     halo potential (right). The models shown are radially 
     and tangentially anisotropic models 
     constructed with the method of G91 (dashed lines), the isotropic
     model (solid), and two  
     Osipkov-Merritt models (dotted lines). Note that while $\beta$ is 
     a function of three-dimensional radius $r$ and $\sigma_{\rm fit}$,
     $h_4$ are 
     observed quantities depending on projected radius $R$, there
     is a close correspondence between features
     in these profiles. See text. }
 \label{models}
\end{figure*}

\section{Velocity profiles in spherical galaxies}

To better understand the relation between \vp--shape, anisotropy, and
gravitational potential, we have constructed a large number of
anisotropic models for spherical galaxies in which the stars follow a
Jaffe (1983) profile.  The gravitational potential was taken to be
either that of the stars (self-consistent case), or one with
everywhere constant rotation speed (`halo potential').  The latter
case corresponds to a mass distribution with a dark halo which has
equal density as the stars at $r\ssimeq 0.4 r_J$, and equal interior
mass at $r\ssimeq r_J$, where $r_J$ is the scale radius of the Jaffe
model.  Anisotropic quasi-separable distribution functions (\df s)
$g(E)\, h(E,L^2)$ were calculated by the method of Gerhard (1991;
G91), but contrary to G93 the circularity function (which specifies
the distribution of angular momenta on energy shells) was allowed to
vary with energy. These models include \df s in which the anisotropy
changes radially, from tangential to radial or vice-versa, from
isotropic to radial to tangential, etc. (see Fig.~\ref{figatwo}). For
comparison, we have also constructed families of Osipkov (1979) -
Merritt (1985) models which become strongly radially anisotropic
beyond a certain radius. Details of this model data base and the
properties of their \vp s are given in Appendix A and in Jeske (1995);
here we only give a brief summary relevant for the comparison with NGC
6703.
 
The shapes of the observable \vp s are most sensitive to the
anisotropy of the \df, but depend also on the potential (G93). For
rapidly falling luminosity profiles, the \vp s are dominated by the
stars at the tangent point. Then radially (tangentially) anisotropic \df s
lead to more peaked (more flat-topped) \vp s than in the isotropic
case; in terms of the Gauss-Hermite parameter $h_4$, this corresponds
to $h_4 \!>\!  (h_4)_{\rm iso}$ and $h_4 \!<\! (h_4)_{\rm iso}$,
respectively (Figs.~8,9 in G93).

Fig.~\ref{models}\ shows that these trends are also seen in the present models in
which both the luminosity density and the anisotropy change with
radius. An increase in radial
(tangential) anisotropy at intrinsic radius $r$ is accompanied by an
increase (decrease) of $h_4$ at projected radius $R\ssimeq r$. The
correspondence is strongest in the models' outer parts, but is also
seen to a lesser extent in the centre of a Jaffe model where
$\rho(r)\propto r^{-2}$ -- contrary to a homogeneous core where
radial orbits lead to broadened \vp s (Dejonghe 1987). Quantitatively
the correspondence depends also on the anisotropy gradient.
Osipkov-Merritt-models show a reversal of this trend near their
anisotropy radius $r_a$ due to the large number of high-energy radial
orbits all turning around near $r_a$; this leads to flat-topped \vp s
in a small radius range near $r_a$. However, the properties of these
models are extreme and they are in general not very useful for
modelling observed \vp s.
 
Fig.~\ref{models}\ and Figs.~8,9 in G93 also show that as the mass of
the model at large $r$ is increased at constant anisotropy, both the
projected dispersion and $h_4$ increase. Increasing $\beta$ at
constant potential, on the other hand, lowers $\sigma$ and increases
$h_4$.  This suggests that by modelling $\sigma$ and $h_4$ both mass
$M(r)$ and anisotropy $\beta(r)$ can in principle be found.

\vfill

\section{Modelling absorption line profile data}

 Having seen the effect of anisotropy and potential variations on the
line profile parameters, we now proceed to construct an algorithm
by which the distribution function and potential of a spherical galaxy
can be constrained from its observed $\sigma$ and $h_4$--profiles.
Such absorption line profile data contain a subset of the information
given by the projected distribution function $N(\b R,\vpar)$, which in
the spherical case is related to the full \df\ by
\leqnam{\projdf}
\begin{eqnarray}
	N(R,\vpar) &=& \int \!dz \int\!\!\int \!dv_x dv_y f(E,L^2) \\
	           &=& \int \!dz \int\!\!\int \!dv_x dv_y
		f[\thalf v_r^2 + \thalf v_t^2 + \Phi(r), r^2 v_t^2].
							\nonumber 
\end{eqnarray}
 Here the position on the sky is specified by $\b R=(x,y)$; $R=|\b
R|$. Velocities in the sky plane are denoted by $(v_x,v_y)$, $z$ and
$\vpar$ are the line-of-sight position and velocity, and $v_r$ and
$v_t$ are the intrinsic radial and tangential velocities. In spherical
symmetry, the \df\ $f(E,L^2)$ is a function of energy and squared
angular momentum only.

Notice from eq.~\projdf) that the projected \df\ depends linearly on
$f$, but non-linearly on the potential $\Phi(r)$.  Thus $f$ will
generally be easier to determine from $N(R,\vpar)$ than $\Phi$.
Moreover, while considerations like those in the last section do
suggest that, in spherical symmetry and for positive $f(E,L^2)$, both
the \df\ {\sl and} the potential can be determined from $N(R,\vpar)$,
there is no theoretical proof that this is in fact true. We only know
that in a {\sl fixed} spherical potential the \df\ is uniquely
determined from $N(R,\vpar)$ (Dejonghe \& Merritt 1992).  For these
reasons we have found it useful to split our problem into two parts
(see also Merritt 1993):

(1) We fix the potential $\Phi$, and from the photometric and
kinematic data determine the ``best'' \df\ $f$ for this
potential. Because in practice the surface brightness (\sb) profile is
much better sampled than the kinematic observations and also has
smaller errors, we treat it separately and determine the stellar
luminosity density $j(r)$ at the beginning. The kinematic data are
then used to determine the ``best'' $f$ for given $j(r)$ and
$\Phi(r)$, by approximately solving equation \projdf) as a linear
integral equation.

(2) We then vary $\Phi$ to find that potential which allows the best
fit overall.  At present, it is not practical in step (2) to attempt
to determine the potential non-parametrically. Rather, we choose
a parametrized form for $\Phi$, and find the region in parameter
space for which the ``best'' \df\ as determined in step (1) reproduces
the data adequately.

In view of the modelling of NGC 6703 in Section 5, we have considered
the following family of potentials, including a luminous and a dark
matter component: The stellar component is approximated as a Jaffe
(1983) sphere, with scale-radius $r_J$ and total mass $M_J$, so that
\leqnam{\phijaffe}
\begin{equation}
\Phi_L(r)  = {GM_J\over r_J} \ln {r\over r+r_J}.
\end{equation}
The dark halo has an asymptotically flat rotation curve,
\leqnam{\vchalo}
\begin{equation}
v_c(r) = v_0 {r\over \sqrt{r^2 + r_c^2}},
\end{equation}
so that its potential is
\leqnam{\phihalo}
\begin{equation}
\Phi_H(r) = \thalf v_0^2 \ln(r^2+r_c^2).
\end{equation}
 This is specified by the asymptotic circular velocity $v_0$ and the
core radius $r_c$. Both the luminous and dark halo components can
be modified when needed, and need not be analytic functions.

In testing our method below, we use parameters adapted to NGC
6703. This galaxy is well-fit by a Jaffe profile (Section 2), so $r_J$
is known. This leaves three free parameters, the mass $M_J$ or
mass--to--light ratio $\Upsilon$ of the stellar component, and the
halo parameters $r_c$ and $v_0$.  If one assumes that the central
kinematics is dominated by the luminous matter, $\Upsilon$ can be
determined. Then only the two halo parameters $r_c$ and $v_0$ are free.
The assumption of maximum stellar $\Upsilon$ is similar to the maximum
disk assumption in spiral galaxies.

In any of the potentials specified by eqns.~\phijaffe)--\phihalo) we
determine the \df\ by the algorithm described in Section 4.1 below.
To assess the significance of the results obtained, we test the
algorithm on Monte Carlo--generated pseudo data in Section 4.2.  For
kinematic data with the spatial extent and observational errors such
as measured for NGC 6703, the algorithm recovers a smooth spherical
\df\ $\sim 70\%$ of the time to an \rms\ level of $\sim 12\%$, taken
inside three times the radius of the outermost kinematic data
point. In Section 4.3, we investigate the degree to which the
gravitational potential can be constrained from similar data.

\subsection{Recovering $f$ from $\sigma$ and $h_4$, given $\Phi$}

As discussed above, the projected distribution function $N(r,\vpar)$
suffices to determine \df\ $f(E,L^2)$ uniquely.  In practice, however,
only incomplete and noisy data are available in place of $N(r,\vpar)$,
and contrary to the two-dimensional function $N(r,\vpar)$, the
observed $\sigma(R_i)$ and $h_4(R_i)$ contain only one-dimensional
information.  This suggests that we can hope to recover only the gross
features of $f$ from such kinematic data. Indeed, the anisotropy parameter
$\beta(r)$ seems to be essentially fixed from accurate $h_4$
measurements (e.g., Figs.~8,9 in G93). Local fluctuations in the
\df\ will be inaccessible, but as we will show, smooth \df s can
be recovered with reasonable accuracy from presently available data.

To solve the inversion problem, we first compute a set of self--consistent
models $f_k(E,x)$ for the stellar density $j(r)$, in the fixed potential
$\Phi(r)$. The $f_k(E,x)$ are models of the kind discussed in Section 2
and Appendix A; $E$ and $x$ are the energy and circularity integrals
of the motion. Then we write the \df\ as a sum over these ``basis''
functions:
\leqnam{\compdf}
\begin{equation}
f=\sum_{k=1,K} a_k f_k(E,x).
\end{equation}
  We do not need to use a doubly infinite, complete set of basis
functions because most of the high--frequency structure represented by
the higher--order basis functions in such a set will be swamped by
noise in the observational data. It is sufficient to choose the number
of basis functions, $K$, and the $f_k(E,x)$ themselves such that the
data can be fitted with a mean $\chi^2 \simeq 1$ per data point.  We
have found it advantageous to use the isotropic model plus 
tangentially anisotropic basis
models, because with these the anisotropy of the final composite \df\
\compdf) can be varied in a more local way than with radially
anisotropic components. Since the $a_k$ can be negative, it is of
course no problem to generate a radially anisotropic \df\ from the
isotropic model plus a set
of tangential basis models.

\begin{figure}
\vspace*{0cm} \centering \getfig{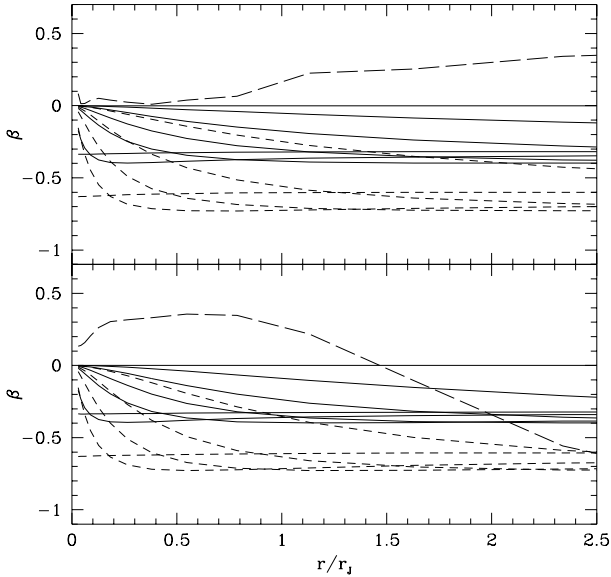} \vspace*{0.cm}
\caption[figbasis]
	{Anisotropy parameter $\beta$ for a subset of basis functions
$f_k$ used in the self--consistent Jaffe potential (top) and in a
mixed Jaffe plus halo model (bottom).  The top full line in each panel
shows the isotropic model. All other basis models are tangentially
anisotropic.  Two values of asymptotic anisotropy as $r \rightarrow
\infty$ are used (full and dashed lines). For illustration, the
long--dashed lines show the $\beta$--profiles of the best--fitting \df s
derived with these bases from the NGC 6703 data, in both potentials.}
\label{figbasis}
\end{figure}

In Fig.~\ref{figbasis} we illustrate the basis models we have used, by
plotting radial profiles of the anisotropy parameter $\beta(r)$ for a
subset of them. For these basis functions $\beta(r)$ can be regarded
as a measure of the extent of the \df\ in circularity $x$, at the
energy $E=\Phi(r)$. The figure shows that the basis resolves three
steps in $x$ in the limits $r\rightarrow 0$ and $r \rightarrow \infty$;
at intermediate energies, it has much finer resolution. The majority
of the number of functions is spent on resolving the energy
dependence, i.e., on placing the main gradient zones of the functions
$f_k$ in the ($E,x)$ plane on a relatively dense grid in energy. The
main difference to using power law components (Fricke 1952, and, e.g.,
Dejonghe \etal 1996) is that our basis models are already reasonable
in the sense that they are viable dynamical models for the density in
question, and that the superposition is used only to match the
kinematics. We have typically used $K\lta 20$ such functions; this
proved sufficient even for analysing pseudo data of much better
quality than we have for NGC 6703.

Each of the $f_k(E,x)$ reproduces the stellar density distribution
$j(r)$; so the $a_k$ satisfy
\leqnam{\norm}
\begin{equation}
\sum_{k=1,K} a_k = 1.
\end{equation}
Moreover, the coefficients $a_k$ must take values such that
\leqnam{\ineqs}
\begin{equation}
\sum_{k=1,K} a_k f_k(E,x) \ge 0
\end{equation}
 everywhere in phase-space. In practice, these positivity constraints
are imposed on a grid in energy and circularity $(E\equal E_i, x\equal
x_j)$.  Subject to these constraints the $a_k$ are to be determined
such that the kinematics predicted from the \df\ \compdf) match the
observed kinematics in a minimum $\chi^2$--sense.

The comparison between model and data is not entirely straightforward,
however.  Since the measured ($v$, $\sigma$, $h_3$, $h_4$) are
obtained by {\sl fitting} to the line profile (the observational
analogue of the projected \df), they unfortunately depend non-linearly
on the galaxy's underlying \df. Thus we cannot use the observed $v$
and $\sigma$ in a linear least squares algorithm to determine the
$a_k$ from the data -- they cannot be written as moments of $f$. In
the fitting process, they have to be replaced by quantities that do
depend linearly on the \df. Moreover, the error bars for these new
quantities generally depend not only on the observed error bars of $v$
and $\sigma$, but also on the errors and the values of the line
profile parameters $h_3$ and $h_4$. They must therefore be determined
with some care.

We have investigated several schemes along these lines. The following
seemed to do best in recovering a known underlying spherical \df\ from
pseudo data.  From the measured ($\sigma, h_4$), we compute an
approximation to the true velocity dispersion (second moment)
$\=\sigma^2(R_i)$, by integrating over the line profile (for negative
$h_4$, until it first becomes negative). We also evaluate a new set of
even Gauss-Hermite moments $s_n(R_i;\s\sigma)$ from the data, using
fixed, fiducial velocity scales $\s\sigma(R_i)$ (G93, Appendix B).
We have found it convenient to take for these $\s\sigma(R_i)$ the
velocity dispersions $\sigma_{\rm iso}(R_i)$ of the isotropic model
with the galaxy's stellar density, in the current potential $\Phi(r)$.

The velocity profile moments of the basis function models are
transformed to the same velocity scales $\s\sigma(R_i)$.  The moments
$\=\sigma^2(R_i)$ and $s_n(R_i;\s\sigma)$ of the composite \df\ then
depend linearly on the corresponding moments of the basis function
models. For a regularized model they are smooth functions of $R$, of
which the (noisy) observational moments are assumed to be a random
realisation within the respective errors.

To determine the best-fitting coefficient $a_k$ we minimize
the sum over data points $i$ of all
\leqnam{\eqsigma}
\begin{equation}
 \chi^2_{\sigma,i} \equiv w^2_\sigma(R_i) \Big[ \=\sigma^2(R_i)
	- \sum_{k=1,K} a_k \=\sigma^2_k(R_i) \Big]^2
\end{equation}
and
\leqnam{\eqsn}
\begin{equation}
 \chi^2_{n,i} \equiv w^2_n(R_i) \Big[  s_n(R_i;\s\sigma)
	- \sum_{k=1,K} a_k s_n^{(k)}(R_i;\s\sigma) \Big]^2
\end{equation}
 for $n=2,4$.  These equations make use of the fact that all the $f_k$
are self-consistent models for the same $j(r)$, so that all surface
density factors $\mu_k(R_i)\equal \mu(R_i)$ cancel.

To determine the weights $w_\sigma$ and $w_n$, we have done Monte
Carlo simulations studying the propagation of the observational errors
$\Delta\sigma$ and $\Delta h_4$.  Based on the results of these
simulations, we have chosen
\begin{eqnarray}
\leqnam{\eqweightsigma}
  w_\sigma^{-1}(R_i)&=&2.4 \=\sigma(R_i) [\=\sigma(R_i)/\sigma(R_i)]
					\Delta\sigma(R_i), \\
\leqnam{\eqweightstwo}
  w_2^{-1}(R_i)     &=&0.7 \Delta\sigma(R_i)/\sigma(R_i), \\
\leqnam{\eqweightsfour}
  w_4^{-1}(R_i)     &=&0.95 \alpha \Delta h_4.
\end{eqnarray}
 The coefficients are representative in the range of values taken by
the observed error bars and the measured $h_4$.  In the presently used
fitting procedure, the additional dependence on $h_4$ and the
respective ``other'' error bar is neglected -- the tests below show
that this leads to satisfactory results. Since to first
order the $s_2$--moment thus measures the shift from $\sigma$ to
$\=\sigma$, the parameter $\alpha$ in eq.~\eqweightsfour) will be near
$\alpha\simeq 2$. We have fixed $\alpha$ by requiring that the
distributions of $\chi^2_\sigma$ and $\chi^2_{h4}$ have equal width
(see below); this results in $\alpha=1.7$.

Finally, we assume that the \df\ underlying the observed kinematics is
smooth to the degree that is compatible with the measured data.
Clearly, unless such an assumption is made, it is impossible to
determine a function of two variables, $f(E,L^2)$, from a small number
of data points with real error bars. One way to ensure that the \df\
is smooth is to use only a small number of terms in the expansion
\compdf). However, this is not a good way of smoothing as it biases
the recovered \df\ towards the functional forms of the few $f_k$ that
are used in the sum. A better way of smoothing is the method of
regularization, as recently discussed by Merritt (1993) in a similar
context. Regularization has tradition in other branches of science,
and different variants exist; for references see Press \etal (1986)
and Merritt's paper (1993). In the algorithm here, regularization is
implemented by taking the number of basis functions large enough
(typically, $16-20$) that the data can be modelled in some detail, and
then constraining the second derivatives of the composite \df\ to be
small. That is we also seek to minimize
\leqnam{\regul}
\begin{eqnarray}
  \Lambda(f)_{ij} \equiv w^2_r(E) \times \hfill 
	\phantom{this is the most stupid tex macro that is !!!} && \\
    \;\;  \left[ \Big( D_E^2 \ppbyd{f}{E^2} \Big)^2 \!+
	2 \Big( D_E \ppbyd{f}{E\p x} \Big)^2 \!+
	\Big( \ppbyd{f}{x^2} \Big)^2 \right]_{E=E_i, x=x_j} && \nonumber
\end{eqnarray}
 on a grid of points ($E_i,x_j$) in the energy and circularity integrals.
Here we normalize to the isotropic \df\ to ensure that fluctuations
in the composite \df\ are penalized equally at all energies; i.e.,
$w_r(E)=1/f_{\rm iso}(E)$. The constant $D_E$ is proportional
to the range in potential energy in the respective model.

To find a regularized spherical \df\ for given kinematic
data in a specified potential, we thus minimize the quantity
\leqnam{\minimize}
\begin{equation}
  \Delta^2 \equiv \sum_{i=1}^I \left\{
		\chi^2_{\sigma,i} + \sum_{n=2,4} \chi^2_{n,i} \right\}
	+ \lambda \sum_{i,j} \Lambda(f)_{ij}
\end{equation}
 for given regularization parameter $\lambda$, subject to the equality
and inequality constraints \norm) and \ineqs).  For the actual
numerical solution we have used the NETLIB routine LSEI by Hanson and
Haskell (1981). Once the best model is found, we redetermine its
quality by evaluating its deviations from the actually measured data:
\leqnam{\eqchisq}
\begin{eqnarray}
\chi^2_\sigma &=& I^{-1} \sum_{i=1}^I \Big[\sigma(R_i) - \sigma^{(f)}(R_i) 
				      \Big]^2 / (\Delta\sigma)^2(R_i), \\
\chi^2_{h4}   &=& I^{-1} \sum_{i=1}^I  \Big[ h_4(R_i) - h_4^{(f)}(R_i) 
				      \Big]^2 / (\Delta h_4)^2(R_i).
\end{eqnarray}
The parameters $\sigma^{(f)}(R_i)$ and $h_4^{(f)}(R_i)$ are determined by
fitting a Gauss--Hermite series to the velocity profiles of the
best fitting model; $I$ is the number of kinematic data points.

\subsection{Tests with model data}

\begin{figure}
\vspace*{0cm}
\centering
\getfig{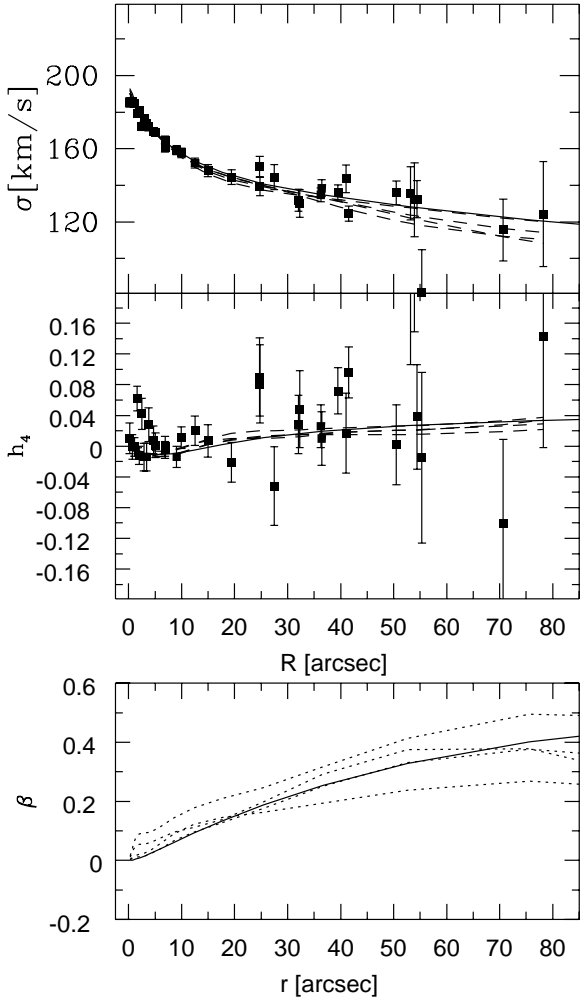}
 \vspace*{0.cm}
 \caption[kinlambdar]
    {Model analysis of pseudo-data generated from a radially
anisotropic model \df\ with the sampling and the measured error bars
of NGC 6703 (see text). The full curves show the true profiles of
projected velocity dispersion, line-of-sight velocity distribution
parameter $h_4$, and anisotropy parameter $\beta$ of the underlying
\df. The dashed and dotted lines show the $\sigma$, $h_4$, and $\beta$
profiles of five regularized composite \df s which were computed by
the method of Sect.~4.1 with $\lambda\equal 10^{-4}$. One of the four
curves in each panel corresponds to the data points actually shown in
this figure; the other three curves derive from statistically identical
kinematic data sets with randomly different values for $\sigma$ and
$h_4$ within the same (Gaussian) errors.
}
\label{kinlambdar}
\vspace*{0cm}
\end{figure}

\begin{figure}
\vspace*{0cm}
\centering
\getfig{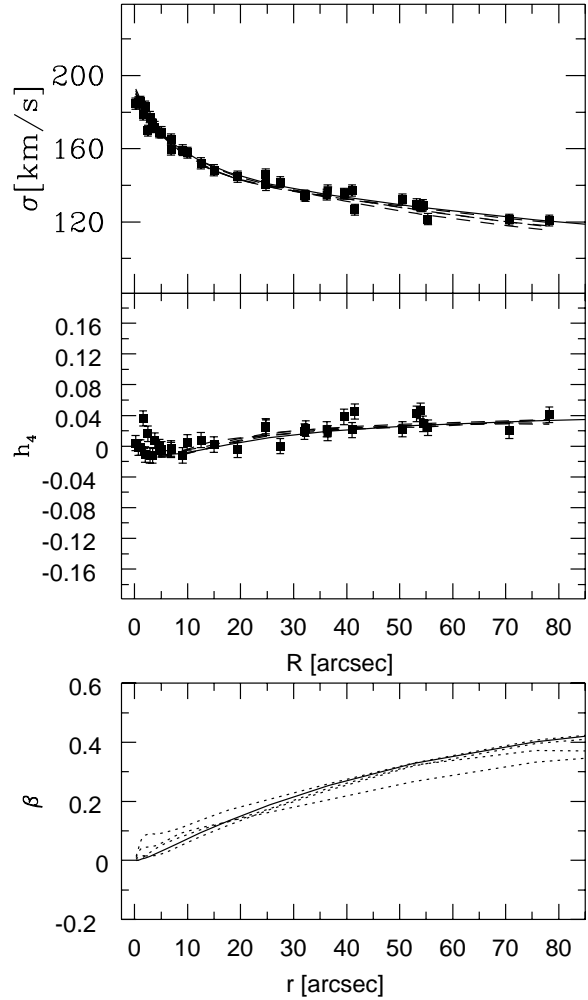}
 \vspace*{0.cm}
 \caption[kinlambdam]
    {Model analysis of pseudo-data generated from the same radially
anisotropic model \df\ as in Fig.~\ref{kinlambdar}\, but assuming
radially constant error bars $\Delta\sigma\equal 3\kms$ and $\Delta
h_4\equal 0.01$.  The full curves again show the true $\sigma$, $h_4$,
and $\beta$ profiles of the underlying \df. The dashed lines show the
projected $\sigma$ and $h_4$ profiles derived by the method of
Sect.~4.1 from the pseudo-data shown, with $\lambda\equal
6\times10^{-5}$.  The results obtained for these projected quantities
from statistically identical data sets are now nearly identical.
The dotted curves show the uncertainties that remain
in the deprojected quantity $\beta$ even with such small error bars.}
 \label{kinlambdam}
\end{figure}

We have tested this method by applying it to kinematic datasets
generated in the following Monte Carlo--like way. First, velocity
dispersions and $h_4$-parameters are calculated from a theoretical
\df\ of specified anisotropy in a known potential, and are
interpolated to the radii $R_i/r_J$ at which observed data points are
available for NGC 6703. Error bars at these $R_i$ are taken to be
either the measured error bars for NGC 6703 (a realistic dataset), or
are taken to be independent of radius with $\Delta\sigma\equal 3\kms$
and $\Delta h_4\equal 0.01$ (an idealized dataset). Then pseudo data
are generated from the model values of $\sigma$ and $h_4$ at $R_i$ by
adding Gaussian random variates with 1-$\sigma$ dispersion
corresponding to the respective $\Delta\sigma$ or $\Delta h_4$ error
bar at this point. Figs.~\ref{kinlambdar}, \ref{kinlambdam} show
datasets generated in this way from a radially anisotropic model in a
potential of a self-consistent Jaffe sphere with $r_J=46.''5$,
$GM_J/r_J= 272\kms$, and a dark halo with $v_0=220\kms$, $r_c=56''$.
This potential was chosen because it lies in the middle of the range
of acceptable potentials for NGC 6703 (see Sect. 5.2), so that any
systematic errors in our analysis will be similar for this model and
for the galaxy itself.

We have used the regularized inversion algorithm described in the last
subsection to analyse several such pseudo data sets, and have
determined composite \df s as a function of the regularization
parameter $\lambda$. The algorithm was given 16 basis function
models. These included the isotropic model and a variety of
tangentially anisotropic models with different anisotropy radii and
circularity functions but not, of course, the radially anisotropic
model from which the data are drawn (see Fig.~\ref{figbasis} and
Appendix A). For each composite model returned by the algorithm we
have determined two diagnostic quantities. The first is the mean
$\chi^2$ per $\sigma$ and $h_4$ data point, $\chi^2_{\sigma+h4}$,
which measures the level at which this model fits the data from which
it was derived. The second is the rms deviation between the returned
\df\ and the true \df\ of the model from which the data were drawn, in
some specified energy range.

The use of these diagnostic quantities requires some further comments.
As usual, the number of degrees of freedom in such a regularized
inversion problem is not well--determined. For near-zero $\lambda$, in
the case at hand we can adjust $16$ coefficients $a_k$ and an overall
mass scale. The total number of $\sigma$ and $h_4$--data points is
$70$; thus in this limit the number of degrees of freedom is $53$. For
large $\lambda$, on the other hand, the recovered \df\ will be linear
in both $E$ and $x$ and the values of the $a_k$ are essentially
fixed. Then the number of degrees of freedom approaches $69$. In both
cases, it is of the order of the number of data points. Hence the use
of $\chi^2_{\sigma+h4}$ per $\sigma$ and $h_4$ data point instead of a
reduced $\chi^2$.

\begin{figure}
\vspace*{0cm}
\centering
\getfig{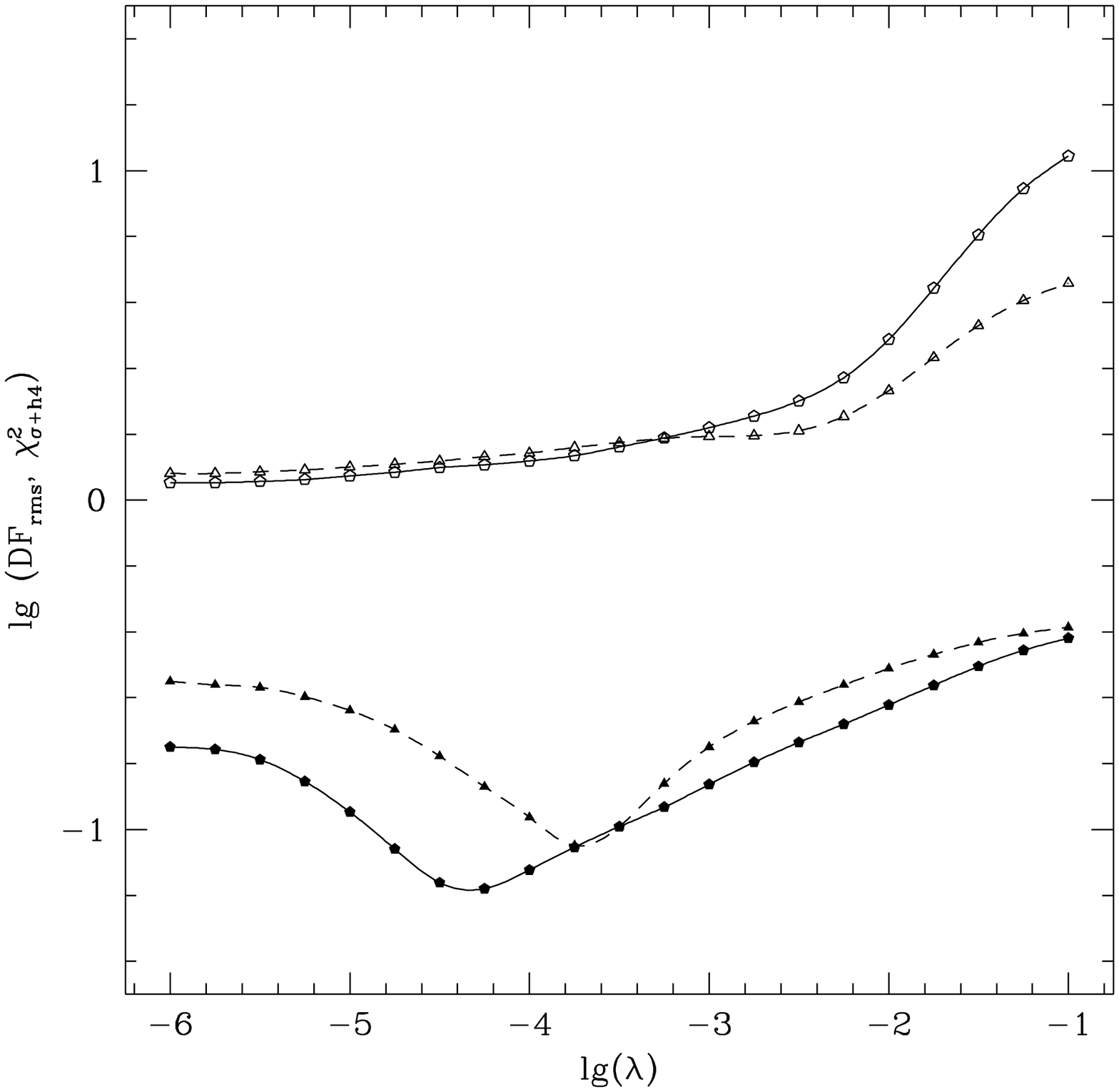}
 \vspace*{0.cm}
 \caption[residulambda]
    {Model results as a function of the regularization parameter
$\lambda$, for the two pseudo-data sets shown in
Figs.~\ref{kinlambdar} and \ref{kinlambdam}. The two data sets are
flagged by triangles (Fig.~\ref{kinlambdar}) and pentagons
(Fig.~\ref{kinlambdam}).  The upper two curves marked by the open
symbols show the total $\chi^2$ per $\sigma$ and $h_4$ data point of
regularized composite models with 16 basis functions. The lower two
curves marked by the filled symbols show the rms deviation between the
recovered composite model \df s and the true \df\ from which the data
sets were generated. This rms deviation was evaluated on a grid in
energy and angular momentum corresponding to radii $\lt 3$ times
the radius of the outermost data point.}
 \label{residulambda}
\end{figure}

\begin{figure}
\vspace*{0cm}
\centering
\getfig{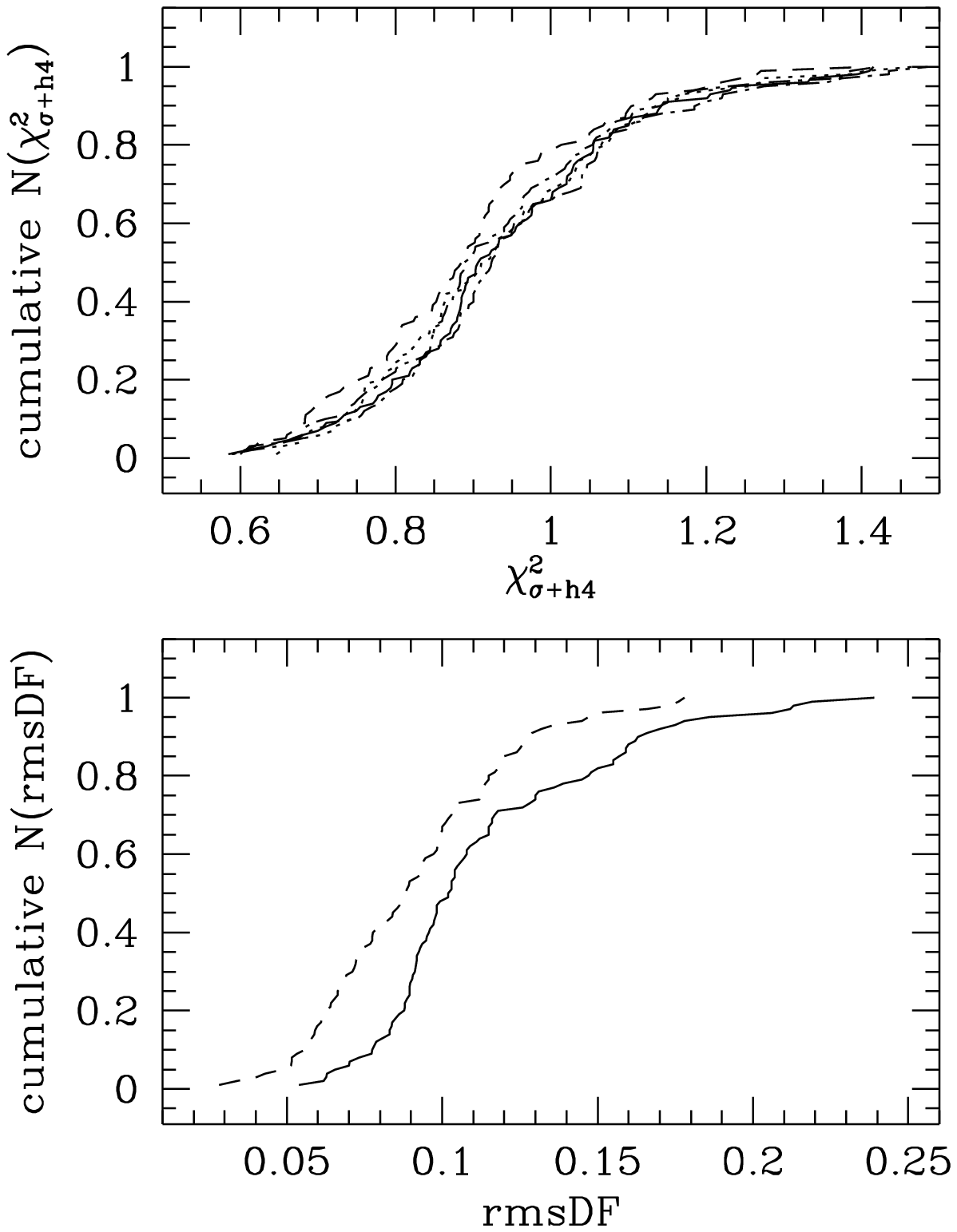}
 \vspace*{0.cm}
 \caption[distrchisq]
    {Top: The cumulative distribution of the total normalized
$\chi^2_{\sigma+h4}$, for dynamical models recovered from 100 random
Gaussian data sets derived from the underlying true \df\ and the
observational errors in Figs.~\ref{kinlambdar} (solid line) and
\ref{kinlambdam} (dashed).  Bottom: Cumulative distribution of
residuals between true and recovered distribution function, evaluated
on a grid extending to three times the radius of the last data point,
for the same 100 datasets from both models. Also shown in the top
panel are the cumulative $\chi^2_{\sigma+h4}$--distributions for data
drawn from radially anisotropic models in the two potentials
corresponding to two of the extreme solid lines in Fig.~\ref{allvc},
with the same error bars as in Fig.~\ref{kinlambdar} (dot--dashed
lines), and for a self--consistent model with more complicated
anisotropy structure (dotted line).}
\label{distrchisq}
\end{figure}

The second diagnostic measuring the accuracy of the recovered \df\
must clearly depend on the range in energy over which it is
calculated.  Typically, a kinematic measurement at projected radius
$R_i$ contains information about the \df\ in a range of energy above
the energy of the circular orbit at radius $R_i$. The precise upper
end of this range is model--specific; it depends on the potential, the
kinematic properties of the \df\ itself, and, through the projection
process, also on the stellar density profile. The use of the outermost
kinematic data points thus contains, explicitly or implicitly,
assumptions on the radial smoothness of these quantities. In a typical
elliptical galaxy problem, the \df\ must be known fairly accurately at
around $3 R_i$ for the projected kinematics at $R_i$ to be securely
predicted, and for radially anisotropic models, even the values of the
\df\ near $10R_i$ can make some difference.  The \rms\ residuals in
the recovered \df\ given below have therefore been calculated in a
range of energies extending from $\Phi(r\equal 0)$ to $\Phi(r\equal
3R_m)$, where the last kinematic data point in the data sets
used is located at $R_m=1.68 r_J$.

Figure~\ref{residulambda}\ shows these quantities as a function of the
regularization parameter $\lambda$ for the two pseudo data sets shown
in Figs.~\ref{kinlambdar}, \ref{kinlambdam}. For small values of
$\lambda$, the composite models fit the data accurately, but the
recovered \df\ is not very accurate because it contains large spurious
oscillations depending on the particular values of the data
points. For large values of $\lambda$, the models are so heavily
smoothed that they neither fit the data well, nor do they represent a
good approximation to the true \df. The optimal regime is where the
smoothing is large enough to damp out the spurious oscillations, but
still permits resolving the important structures in the underlying
\df. For values of $\lambda$ in this regime the fit to the data is
still satisfactory, and the representation of the \df\ is optimal.
Fig.~\ref{residulambda} shows that the \rms\ residuals of the \df\ go
through minima at values of $\lambda\simeq2\times10^{-4}$ for the pseudo
data in Fig.~\ref{kinlambdar} and $\lambda\simeq5\times10^{-5}$ for those
in Fig.~\ref{kinlambdam} [when $D_E=1$; see below eq.~\regul)].

The shapes of the $\chi^2_{\sigma+h4}(\lambda)$ curves were found to
be always similar to the upper curves in Fig.~\ref{residulambda}.  The
shapes of the corresponding lower curves in Fig.~\ref{residulambda}\
are more variable.  The resulting optimal values for $\lambda$ can
vary, depending on the random realisation of the data within the
assumed Gaussian errors, as well as on the distribution function and
potential from which the model values are drawn. We have therefore
investigated 100 realisations of data generated from each of several
model \df s and potentials. Based on these experiments we have fixed
optimal values of $\lambda=1\times10^{-4}$ and $\lambda=6\times
10^{-5}$ for data with error bars like those in
Figs.~\ref{kinlambdar}\ and \ref{kinlambdam}, respectively, for all
models that include a dark halo component.

We have done similar experiments with a self-consistent model that
is radially anisotropic near the center and tangentially anisotropic
in its outer parts, such as might be relevant in tests for dark matter
at large radii. In these tests we have used 20 basis functions.  To
match the corresponding pseudo-data with similar accuracy requires
smaller optimal values of $\lambda$ ($\sim 3\times 10^{-9}$), so as to
compensate for the larger derivatives in eq.~\regul).

Fig.~\ref{distrchisq}\ shows the cumulative distributions of
$\chi^2_{\sigma+h4}$ and of the \rms\ residual in the \df, as
described, for the dynamical models recovered from several such sets
of Monte Carlo data. From the top panel it is seen that our fitting
algorithm will match kinematic data in a known potential with
$\chi^2_{\sigma+h4}\le 1$ about $60-70\%$ of the time, and with
$\chi^2_{\sigma+h4}\gt 1.28$ only $<5\%$ of the time. These numbers
are similar to those expected from a $\chi^2$--distribution with $70$
degrees of freedom (for Gaussian data), which should describe the
statistical deviations of the Monte Carlo data points from the
underlying true \df.  If in modelling the data for NGC 6703 a level of
$\chi^2_{\sigma+h4}\lt 1.28$ cannot be reached, the assumed potential
is not correct with $95\%$ confidence.

The bottom panel of Fig.~\ref{distrchisq}\ shows the distribution of
residuals in the recovered \df\ for one model.  There are three
factors which limit the degree to which a given underlying \df\ can be
recovered. The first is determined by the data, i.e., the size of the
error bars, the sampling, the fact that there are no measurements at
large radii, etc. The second is the level of detail that can be
resolved by the modelling, given the finite number and the particular
form of the basis functions. The third is the amount of small--scale
gradients in the model itself.  Figs.~\ref{residulambda} and
\ref{distrchisq} show that, if the potential is known, data like those
for NGC 6703 (Fig.~\ref{kinlambdar}) allow recovering reasonably
smooth \df s to an \rms\ level of $\lta 12\%$ inside $3R_m$ about
$60-70\%$ of the time. Data with much smaller error bars but the same
sampling (Fig.~\ref{kinlambdam}) would give an \rms\ level of $\lta
10\%$ ($\lta 8\%$ for the other two halo models shown in the top panel
of Fig.~\ref{distrchisq}). In the radial--tangential self--consistent
model the \df\ to $3 R_m$ is recovered with an \rms\ accuracy of
$16\%$ and $35\%$ for data with error bars like those in
Figs.~\ref{kinlambdam} and \ref{kinlambdar}, respectively.  These
comparisons and the results of Fig.~\ref{distrchisq} show that the
former values are dominated by the measurement uncertainties rather
than the resolution in the modelling. This conclusion would be
different for highly corrugated true \df s. However, we would not be
able to recover such \df s from realistic data in any case.

The kinematics of the regularized composite \df s derived from
pseudo-data with the chosen optimal $\lambda$ values are shown in
Figs.~\ref{kinlambdar} and \ref{kinlambdam}.  (For brevity, such model
\df s obtained with near--optimal $\lambda$ will henceforth be denoted
as ``best--estimate models''.)  The good match to the data points is
apparent. The differences in the intrinsic anisotropy parameter
$\beta$ between the recovered models and the true model are
larger. For the data set generated with the observed error bars of NGC
6703 the recovered $\beta$ values are uncertain by $\Delta\beta\simeq
\pm 0.1$, and slightly more at the largest radii where the
observational errors are large. For the pseudo-data with the small
error bars in Fig.~\ref{kinlambdam}, this uncertainty is reduced.

\subsection{Constraining $\Phi$}

\def\chish{\chi^2_{\sigma+h4}}
\def\chis{\chi^2_{\sigma}}
\def\chih{\chi^2_{h4}}

So far we have shown how the stellar \df\ can be recovered from
\vp--data in a known spherical potential. In this Section we return to
the discussion of Section 3 and investigate the degree to which the
gravitational potential itself can be constrained. Eventually, it will
clearly be important to answer the theoretical question of whether, in
principle, the gravitational potential is nearly or even uniquely
determined from the projected \df\ $N(r,\vpar)$. However, we shall not
attempt to do this here; previous theoretical work suggests that the
range of potentials consistent with ideal data is small (G93, Merritt
1993, see Section 3). More relevant at the moment perhaps is the more
practical question of how well the potential can be constrained from
real observational data with realistic error bars, finite radial
extent, and limited sampling. What is the range of pairs $(f,\Phi)$
that correspond to the same data? How does this range shrink as the
data improve?

Here we investigate these questions with a view to the analysis of the
NGC 6703 data below. We again use the model underlying
Figs.~\ref{kinlambdar}, \ref{kinlambdam}; this is a radially
anisotropic \df\ in the potential of a self--consistent Jaffe sphere
and a dark halo with parameters $GM_J/r_J=272\kms$, $r_c/r_J=1.2$ and
$v_0^2/(GM_J/r_J)=0.808$. As before, random Gaussian data sets were
generated from this model, with the positions and error bars of the
data points (i) as in Fig.~\ref{kinlambdar}, (ii) as in
Fig.~\ref{kinlambdam}.  The last data point is at $R_m=1.68 r_J$, as
for the NGC 6703 data, well beyond two effective radii.  However,
contrary to Section 4.2, the data sets used in the following were
specially selected such that $\chis\simeq 1$ and $\chih\simeq 1$.  (It
turns out that the data points shown in Fig.~\ref{kinlambdar} have
less than $5\%$ probability according to Fig.~\ref{distrchisq}.).

For these pseudo data we have determined best--estimate \df s in a
number of assumed potentials, including the underlying true potential.
The potentials were chosen such that (i) they correspond to
approximately constant mass--to--light ratio for $r\ll r_J$, and (ii)
they form a sequence of varying true circular velocity $v_c(R_m)$ at
the radius of the last data point. The sequence, with the slightly
different velocity normalisations appropriate for NGC 6703, is shown
in Fig.~\ref{allvc}. For each potential, specified by the selected
values of the parameters $r_c/r_J$ and $v_0^2/(GM_J/r_J)$, we determined
the goodness--of--fit $\chish$ as a function of the velocity scale
$GM_J/r_J$ from the corresponding best--estimate models. Finally, we
computed the $\chish$ of the best--estimate \df\ for the optimum
velocity scale.  This optimal velocity scale usually turns out
slightly different for the two pseudo data sets.

\begin{figure}
\vspace*{0cm}
\centering
\getfig{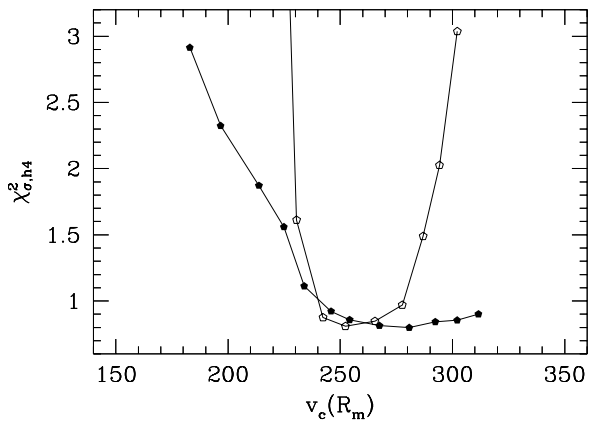}
 \vspace*{0.cm}
 \caption[chisqvcpot.eps]{Goodness of fit $\chish$ for best--estimate
	\df s derived in a sequence of 
	luminous plus dark matter potentials. The sequence of potentials,
	with rotation curves similar to those shown in Fig.~\ref{allvc},
	is here parametrized by the total circular velocity at the last 
	observational radius, $v_c(R_m)$.  The \df s were fitted
	to pseudo data generated from a model with $v_c(R_m)=242\kms$.
	Solid pentagons: data points with error bars as in
	Fig.~\ref{kinlambdar}. Open pentagons: data points with error 
	bars as in Fig.~\ref{kinlambdam}. }
 \label{chisqvcpot}
\end{figure}

\begin{figure}
\vspace*{0cm}
\centering
\getfig{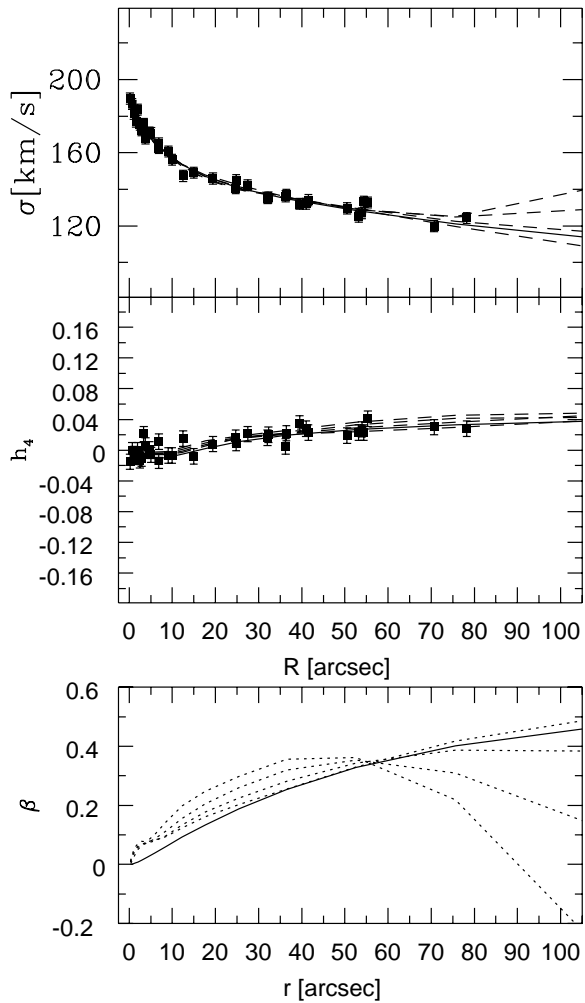}
 \vspace*{0.cm}
 \caption[fits_235m.eps]{Model fits to pseudo velocity dispersion
	and $h_4$ data with ``idealized'' error bars (see text, and
Fig.~\ref{kinlambdam}).  The full line shows the kinematics of the
model from which these pseudo data were generated. The dashed lines
show the kinematics of best--estimate \df s in a sequence of
potentials with increasing halo circular velocity, all of which fit
the data perfectly with $\chish\simeq 1$. The intrinsic anisotropy of
these models (dotted lines) and of the true model are shown in the
bottom panel. }
\label{fits_235m}
\end{figure}

Fig.~\ref{chisqvcpot} shows the goodness--of--fit $\chish$ of the
optimal sequence of best--estimate models as a function of $v_c(R_m)$,
as determined for both model datasets. The underlying true potential
has $v_c(R_m) \equal 242\kms$ with the adopted
parameters. Fig.~\ref{chisqvcpot} shows that only potentials with
$v_c(R_m) < 230\kms$ can be ruled out (have less than $5\%$
probability according to Fig.~\ref{distrchisq}) from the data with
``realistic'' error bars. This conclusion is not surprising given how
large these error bars are. However, the surprise is that also with
``idealized'' data there remains a range of potentials with larger
$v_c(R_m)$ than the true value, in which these data can be fit no
worse than in the true potential: formally, $235 \kms < v_c(R_m) <
285\kms$ with $95\%$ confidence.

Fig.~\ref{fits_235m} shows the predicted kinematics and the intrinsic
anisotropy of the four models in Fig.~\ref{chisqvcpot} that match the
``idealized'' data set with $\chish\simeq 1$. The best--estimate \df s
in the potentials with higher $v_c(R_m)$ than that of the true
potential achieve their good fit to the data by the following means:
Inside $r_J$ they compensate the potential's higher circular speed by
a larger radial anisotropy, which leads to slightly larger
$h_4$--values (cf.~Section~3).  This effect is too small to be
detected even with the small error bars.  Outside $r_J$, they
compensate by a radially increasing tangential anisotropy. In this way,
the velocity dispersions near the edge of the model can be lowered,
because the number of high--energy orbits coming in from outside is
reduced.  Such orbits would contribute large line--of--sight
velocities near their pericentres. The $h_4$ values in the region
concerned are also only barely affected, because the extra tangential
orbits near radius $r$ and the lack of higher energy radial orbits
from beyond $r$ nearly compensate.

Clearly, this mechanism will work less well both when the difference
between the true and attempted circular velocity curves increases, and
when the radial extent, sampling, and quality of the data points
improve. With data yet better and more extended than those in
Fig.~\ref{kinlambdam}, some of the potentials consistent with the
presently used model data could probably be ruled out. Thus it appears
that, if the asymptotic circular velocity is constant, then the value
of that constant can be determined accurately with sufficiently high
quality data.

On the other hand, the potentials we have used are very simply
parametrized functions.  There might well exist more complicated
potential functions, whose circular velocity curves differ in only a
restricted range of radii, that would be impossible to distinguish
even with {\sl extremely} good data. To test this, we have constructed
a truly idealized data set of two times seventy data points with
``idealized'' small error bars as in Fig.~\ref{kinlambdam}, evenly
spaced in radius, and extending to $6r_J$. A model with potential
differing only in the halo core radius (66\% of the true value) 
was found to fit even these
data with $\chish = 0.97$, while for a model with different halo core
radius {\sl and} different asymptotic circular velocity (by $30\kms$)
a satisfactory fit could not be found.

We draw the following conclusions from these experiments:

(i) Velocity profile data with presently achievable error bars contain
useful information on the gravitational potentials of elliptical
galaxies.  In particular, constant--M/L models are relatively easy to
rule out once the data extend beyond $2R_e$. The examples that we have
studied in detail, tuned to the NGC 6703 data, certainly belong to the
less favourable cases, because the dispersion profile is falling.

(ii) The detailed form of the true circular velocity curve is much
harder to determine. Conspiracies in the \df\ are possible that
minimize the measurable changes in the line profile parameters. A good
way to parametrize the results is in terms of the circular velocity at
the radius of the outermost data point. With presently available data,
this can be determined to a precision of about $\pm 50\kms$.

(iii) Better results can be expected from higher--quality data, of the
sort one could expect from the new class of 10m telescopes.  However,
even with such data some uncertainty on the detailed circular velocity
curve will remain -- regardless of whether or not in theory the
potential is uniquely determined from the projected \df\ $N(r,\vpar)$.
Therefore, the combination of the type of analysis presented here with
other information (e.g., from X-ray data) will give the most
powerful results.

\section{The anisotropy and mass distribution of NGC 6703}

\subsection{Constant--M/L model fits}

\def\chih{$\chi^2_{h_4}$}
\def\chis{$\chi^2_\sigma$}
\def\chihm{\chi^2_{h_4}}
\def\chism{\chi^2_\sigma}
\def\chidiagram{\chih-\chis-diagram}

The \sb-profile of NGC 6703 is well fitted by a Jaffe model. The
largest local residuals are $\lta 15\%$ around $R\simeq 25''$, $\lta
10\%$ around $R\simeq 40''$, and smaller elsewhere, in particular at
$R\gt 60''$. A non--parametric inversion of the the surface brightness
profile showed that the deviations from a Jaffe density law are not
significantly larger than those quoted in \sb. 
In the curve of growth, measuring the total luminosity
inside $R$, the residuals are everywhere less than $\sim 2\%$
(Fig.~\ref{obsvns}).  Since the potential $\Phi(r)$ is determined by
the total mass $M(<r)$, we can thus to good approximation use a Jaffe
model for the gravitational potential of the stars. This enables us to
also compare our kinematic data with a large set of self-consistent
dynamical models from Jeske (1995).

We have fitted all models from this database to the observational data
for NGC 6703, taking the fitted Jaffe radius $r_J\equal 46.''5$. All
data points from Fig.~\ref{obsvns}\ were included and weighted equally
with their individual error bars, except for two adjustments. (i) The
error bars of the three $h_4$-points near 25'' have been set equal to
their standard deviation, and (ii) the error bars of the innermost
eleven $h_4$-points have been set to twice the measured values.  These
modifications prevent the total \chih\ to be dominated by these points
which are of no consequence for the halo of NGC 6703. Moreover,
systematic effects may play a role for the innermost data points
(Sect.~2).


\begin{figure}
\vspace*{0cm}
\getfig{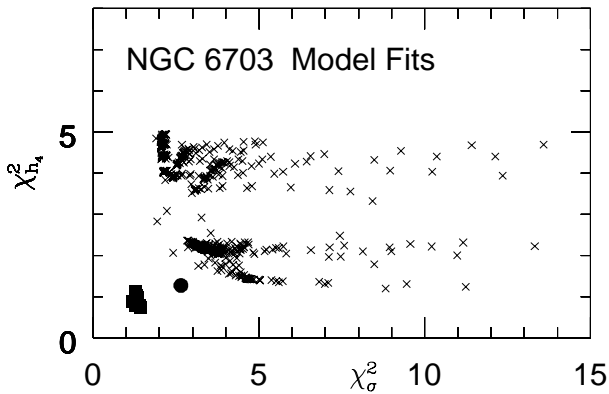}
 \vspace*{0.0cm}
 \caption[\chih-\chis-diagram]
    {\chih-\chis-diagram from fitting dynamical models to the
    kinematic data for NGC 6703. The crosses show fit results for a
    variety of anisotropic, self--consistent models of a Jaffe sphere,
    whose density profile is a good approximation for the luminous
    matter in NGC 6703. They fall upwards and to the right of a curved envelope
    that separates them clearly from a perfect fit, showing that no
    self--consistent model can simultaneously fit both the dispersion
    and line profile shape data. The heavy dot is the best--estimate
    self--consistent model in the stars--only potential, obtained with
    the method of Section 4. The squares show a number of dynamical
    models with a dark halo; these are consistent with the data.}
 \label{chichi}
\end{figure}
 
The velocity scale of each model is matched to optimize the fit to the
$\sigma$-profile; that to the $h_4$-profile is already fixed with no
extra assumption. The resulting values of \chih, \chis\ are normalized
by the number of fitted data points.  In Fig.~\ref{chichi} we plot a
\chidiagram\ for all self-consistent models from Jeske (1995). The
$\chi^2$ values are again normalized to the number of data
points. Also plotted on the figure are best-estimate models in a
number of potentials, constructed with the technique described in
Section 4.  The optimized self-consistent model (the heavy dot) lies
on the bounding envelope of the self-consistent models; it has
$\chism\equal 2.6$, $\chihm\equal 1.3$. The squares show the
normalized $\chi^2$-values for several models with a dark halo whose
rotation curves are shown by the full lines in Fig.~9.  For most of
these $\chism\simeq1.3$, $\chihm\simeq 0.8$.

Fig.~\ref{chichi} shows that {\sl no self-consistent model will fit
the data}: all self-consistent models are clearly separated by a
curved envelope from the lower left hand corner of the diagram, which
corresponds to a perfect fit. Either the velocity dispersion profile
is matched reasonably well, but then the line profiles cannot be
reproduced, or, when the $h_4$-profile is fitted accurately, the
dispersion profile is poorly matched.  The cure for the discrepancy is
to raise {\sl both} $\sigma$ and $h_4$ at large radii.  Thus,
according to Sect.~3 above, we require extra mass at large $r$.  NGC
6703 must have a dark halo.

\subsection{Dynamical models with dark halo}

We will now derive constraints on the gravitational potential of NGC
6703 within the framework of the parametric mass model of
eqs.~\phijaffe)--\phihalo). In doing this we have in mind the
following working picture: The stellar component is assigned a
constant mass-to-light ratio $\Upsilon$, chosen maximally such that
the stars contribute as much mass in the center as is consistent with
the kinematic data. The model for the halo incorporates a constant
density core, and its parameters are chosen such the halo adds mass
mainly in the outer parts of the galaxy if that is necessary. This is
similar to the maximum disk hypothesis in the analysis of disk galaxy
rotation curves.  Within this framework we can determine the maximum
stellar mass-to-light ratio, ask whether it is reasonable, and
constrain the halo parameters.

\begin{figure}
\vspace*{0cm}
\centering
\getfig{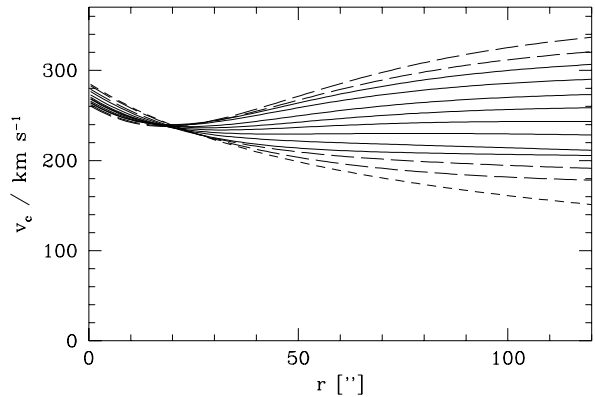}
 \vspace*{0.cm}
 \caption[allvc.eps]{Rotation curves for a sequence of gravitational
	potentials (stars plus dark halo) used in the analysis of NGC 6703.
	The full lines show rotation curves that are consistent with
	the NGC 6703 kinematic data inside the $95\%$ confidence boundary
	at $\lambda=10^{-5}$ (open symbols
	in Fig.~\ref{chisqvc}). The other line styles show rotation curves
	inconsistent with the data; among these is the constant--M/L
	model with no dark halo (short--dashed).}
 \label{allvc}
\end{figure}

Because determining the two potential parameters $r_c$ and $v_0$
together with the model velocity scale (equivalent to the stellar
mass--to--light ratio) is a three--dimensional problem, we will
proceed in steps.  Fig.~\ref{allvc} shows circular rotation curves for
the first sequence of mass models for NGC 6703 that we have
investigated.  In all these models the halo contribution becomes
significant only outside $20''\lta R_e$. The corresponding halo core
radii mostly lie between $1.2r_J$ and $1.7r_J$, i.e., $55''$ and
$80''$.  These values are relatively large because of the falling
dispersion curve in NGC 6703. This implies that we can reliably
determine only one of the halo parameters for this galaxy.

We have chosen the circular velocity $v_c(R_m)$ at the radius $R_m$ of
the last kinematic data point as this parameter.  The sequence in
Fig.~\ref{allvc} was constructed such as to vary $v_c(R_m)$ and the
rotation curve outside $\sim R_e$, while leaving the central rotation
curve nearly invariant.  However, when the model velocity scales are
optimized in the determination of the best--estimate \df s, the
optimal velocity scale is found to correlate inversely with
$v_c(R_m)$.  The rotation curves in Fig.~\ref{allvc} are plotted with
their optimal scaling; then the central rotation curves are no longer
identical, but become scaled versions of each other.

In each of the corresponding potentials, we have constructed the
best--estimate \df\ with optimal velocity scaling, as described in
Section 4. We first fit a composite \df\ to the velocity dispersion
and line profile shape data for a series of values of the unknown
velocity scale.  From this sequence of models, we determine the
optimal value of the model's velocity scale in this potential, and
then recompute the best--estimate model with this velocity scale. In
the following, when speaking of the best--estimate \df\ for a given
potential, we will always imply that the velocity scale has been
optimized in this way.  In the fitting procedure we have used a
regularization parameter $\lambda=10^{-4}$; this was found appropriate
in Section 4.2 for the error bars and sampling of the NGC 6703 data.
Compared with the tests in Section 4, we have included a few extra
basis functions (total $K\equal 20$) to resolve the (possibly not
real, cf.~Section 2) high--frequency structure in the center of NGC
6703.

\begin{figure}
\vspace*{0cm}
\centering
\getfig{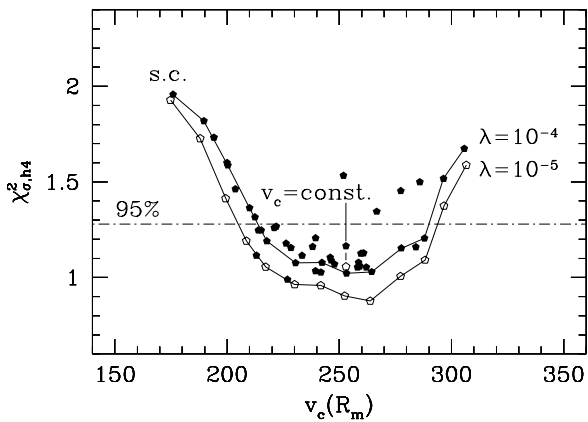}
 \vspace*{0.cm}
 \caption[chisqvc]{Quality with which the kinematics of NGC 6703
can be fitted in different potentials. The figure shows the average
$\chi^2$ per $\sigma$-- and $h_4$ data point, of the best--estimate
distribution function fitted to the velocity dispersion and line
profile data, as a function of the assumed potential's circular
rotation velocity at the observed radius of the last kinematic data
point. Filled symbols show best--estimate models derived with the
optimal $\lambda\equal 10^{-4}$ determined in Section 4.2; open
symbols represent models derived with $\lambda\equal 10^{-5}$.
The self--consistent ($M/L\equal {\rm const.}$) and the $v_c\equal
{\rm const.}$ models are marked separately. The horizontal dashed
line shows the $95 \%$ confidence boundary derived from
Fig.~\ref{distrchisq}. }
 \label{chisqvc}
\end{figure}

Fig.~\ref{chisqvc} shows, as a function of $v_c(R_m)$, the average
$\chi^2$ per $\sigma$-- and $h_4$ data point of the respective \df s
so obtained. The connected solid symbols represent the sequence of
potentials corresponding to Fig.~\ref{allvc}.  The potential with
constant mass--to--light ratio appears in the upper left--hand corner
in Fig.~\ref{chisqvc}; it is inconsistent with the data by a large
margin even for optimum velocity scale (see
Fig.~\ref{distrchisq}). The best--fitting potential with a completely
flat rotation curve has an optimal value of $v_c\equal v_c(R_m)\equal
254\kms$ and $\chish\equal 1.17$. Thus it does not provide the best
possible fit but is consistent with the data.  Of the stars plus dark
halo models illustrated in Fig.~\ref{allvc}, those models in the
sequence with $v_c(R_m) = 210-285 \kms$ have $\chish < 1.28$. This is
consistent with the results of Section 4.3, from which we would expect
that the NGC 6703 data can be fit by a range of gravitational
potentials. Models in the sequence outside this range of $v_c(R_m)$
are inconsistent with the kinematic data at the $95\%$ confidence
level (cf.\ Fig.~\ref{distrchisq}).

\begin{figure}
\vspace*{0cm}
\centering
\getfig{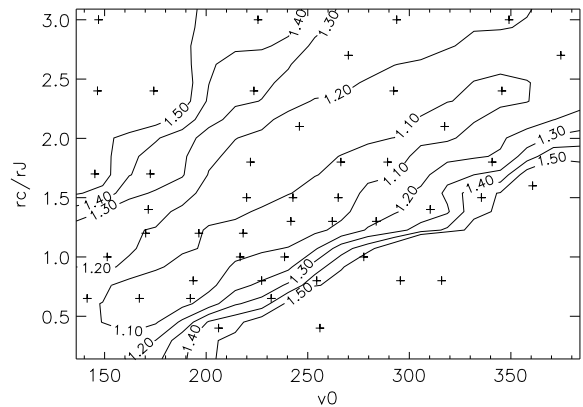}
 \vspace*{0.cm}
 \caption[allphipar]{The $(v_0, r_c)$ halo parameter plane. Values
	of $r_c$ are scaled with respect to $r_J=46''.5$. The luminous
	plus dark matter models investigated are shown as plus signs.
	The contours show lines of constant $\chish$ obtained by
	interpolating between the model values. Acceptable potentials
	lie in a band extending from low $v_0$ and low $r_c$ to high
	$v_0$ and high $r_c$. Models at the upper left are ruled out
	because they do not contain enough mass at large radii. Models
	at the lower right are ruled out because no satisfactory fit
	can be found for any constant value of the stellar mass--to--light
	ratio.}
 \label{allphipar}
\end{figure}

In a second step we have analyzed a more complete set of potentials in
a suitable part of the $(v_0,r_c)$--plane (Fig.~\ref{allphipar}). The
fitting results in these potentials are shown by the isolated filled
symbols in Fig.~\ref{chisqvc}. Combining with the previous results
these allow us to investigate the range of acceptable potentials for
NGC 6703 more fully than with the one-dimensional sequence in
Fig.~\ref{allvc}.

Fig.~\ref{allphipar} shows contours of constant $\chish$ in the
$(v_0,r_c)$--plane.  The most probable potentials lie in a band
extending from low $v_0$ and low $r_c$ to high $v_0$ and high $r_c$.
Thus, as already discussed above, it is not possible to determine both
halo parameters in the NGC 6703 case. However, potentials in the band
of most probable $(v_0,r_c)$ all have circular velocities $v_c(R_m)$
in the same range of $250\pm 40\kms$ as before.

In fact, the best--fitting velocity scales of all models in
Fig.~\ref{allphipar} turn out such that the resulting values of
$v_c(R_m)$ are in the range $[189, 318] \kms$. The fitting procedure
tends to move $v_c(R_m)$ into the correct range even when no
satisfactory \df\ can be found. E.g., some models in the lower right
of Fig.~\ref{allphipar} with $\chish\simeq 1.5$ appear in
Fig.~\ref{chisqvc} at $v_c(R_m)=250-300\kms$, some with $\chish\gta
2.4$ (not shown) at $v_c(R_m)=280-340\kms$.  All models shown in the
upper left of Fig.~\ref{allphipar} fall near the line defined by the
sequence of models discussed at the beginning of this section.  The
fact that $v_c(R_m)$ varies relatively little for a wide range of
luminous matter plus dark halo models suggests that our results, when
expressed in terms of this parameter, are not sensitive to the choice
of halo model in eqs.~\vchalo), \phihalo).

From Fig.~\ref{chisqvc} we conclude that the true circular velocity of
NGC 6703 at 78'' is $v_c(R_m)=250\pm 40\kms$ (the formal $95\%$
confidence interval obtained from the filled symbols, according to
Fig.~\ref{distrchisq}).  In Section 4.3 we found, however, that most
of this indeterminacy is towards large circular velocities, at least
for a galaxy with a dispersion curve like that of NGC 6703. By
contrast, in potentials with lower values of $v_c(R_m)$ than the true
circular velocity, it quickly becomes impossible to find a
satisfactory \df. Based on these results, the lower values in the quoted
range of $v_c(R_m)=250\pm 40\kms$ appear to be the more probable ones.

The open symbols in Fig.~\ref{chisqvc} show that the range of
potentials consistent with the data is enlarged only slightly when the
\df\ is allowed to be less smooth. These points are from
best--estimate models derived in the sequence of potentials of
Fig.~\ref{allvc} with $\lambda=10^{-5}$ instead of the optimal
$\lambda=10^{-4}$. The resulting curve which connects the
$\lambda=10^{-5}$ models in Fig.~\ref{chisqvc} surrounds the
corresponding $\lambda=10^{-4}$ curve.  Generally it appeared that, in
the models obtained with $\lambda=10^{-5}$, the \df\ came close to
zero more easily and more often. The last two facts, when taken
together, suggest that, in addition to the data themselves, the
positivity constraints on the \df\ play an important role in
determining the boundary of the region in $(f,\Phi)$ space that is
consistent with given kinematic data.

\begin{figure}
\centering
\getfig{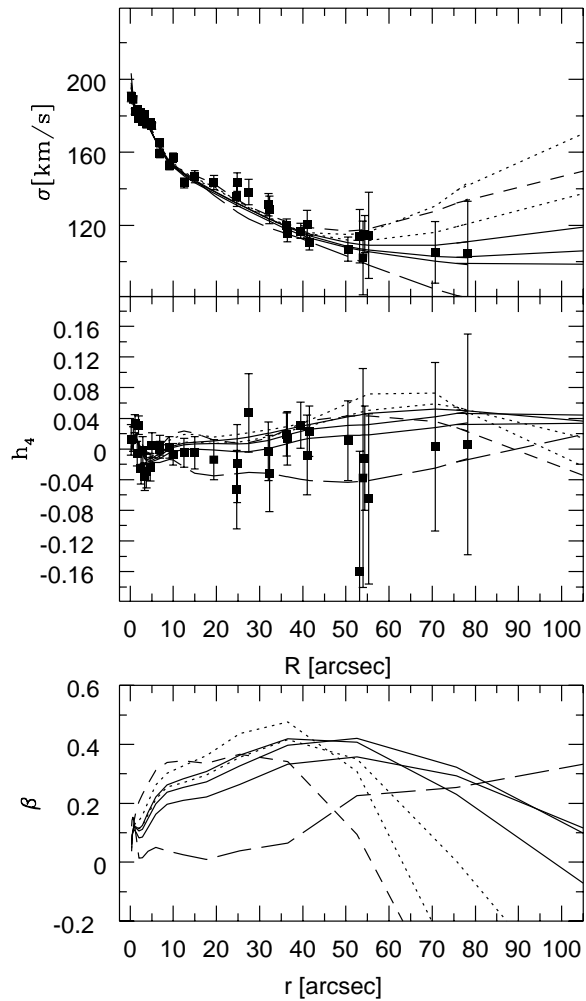}
 \vspace*{-0.4cm}
 \caption[Dynamical models for the kinematics of NGC 6703]
    {Dynamical models for the kinematics of NGC 6703 in several
     luminous plus dark matter potentials, compared to projected 
     velocity dispersion (top panel) and \vp-shape parameter 
     $h_4$ (middle panel). The bottom panel shows the models' intrinsic 
     anisotropy parameter $\beta(r)$, with the same linestyles:
     self--consistent model (stars only; long--dashed), $v_c\equal {\rm const.}$
     model (short--dashed), three models with $v_c(78'')$ in the lower
     part of the acceptable range (full), and two models with $v_c(78'')$
     in the upper part of this range (dotted lines).} 
 \label{kinematicfits}
\end{figure}

Fig.~\ref{kinematicfits} shows best--estimate models in some of the
potentials consistent with the kinematic data (at $\lambda=10^{-4}$).
The three full lines are models with $v_c(R_m) = (218, 231, 242)
\kms$, the two dotted lines are models with $v_c(R_m) = (253, 277)
\kms$.  Also shown is the best--estimate model in the potential of
only the stars with constant M/L (long--dashed), and that in the best
potential with $v_c={\rm const.}=254\kms$ (short--dashed).

The dispersion profile of the best--estimate self-consistent model
falls clearly below the data at both intermediate and large radii, as
does its $h_4$-profile. From the discussion in Section 3, this is a
clear sign of extra mass at large radii. The models including dark
halo contributions mainly differ in the outermost parts of the
velocity dispersion profile. As expected, those with the highest
velocity dispersions at large radii correspond to the potentials with
the largest asymptotic circular velocities.  This again suggests that
with smaller error bars at large radii and with spatially more
extended data, we will be able to significantly narrow down the
uncertainties in the halo parameters. The model with everywhere
constant rotation speed is constrained tightly by the kinematic data
in the central parts, where the $\rho\propto\sim r^{-2}$ profile is
presumably dominated by the stars. It then has some difficulties both
with the $h_4$--values at intermediate radii and the velocity
dispersions at large radii.

The lower part of Fig.~\ref{kinematicfits} shows that, in order to
match the observed kinematics in a potential with large circular
velocity, the \df\ must become rapidly tangentially anisotropic at the
radii of the last data points and beyond. As discussed in Section 4,
this is different from the better known effect of increasing the
projected dispersions in a potential with not enough mass at large
radii, by making the \df\ more tangentially anisotropic. Here, without
the tangential anisotropy, our models would predict too high values of
the velocity dispersion, because too much mass at large radii is
implied. The tangential anisotropy at radii outside those reached by
the observations reduces the number of orbits that come into the
observed range, orbits which would contribute relatively large
line-of-sight velocities near their turning points. In this way the
velocity dispersions can be reduced down to the observed values.
Clearly, more spatially extended data would reduce this freedom.

\begin{figure}
\centering
\getfig{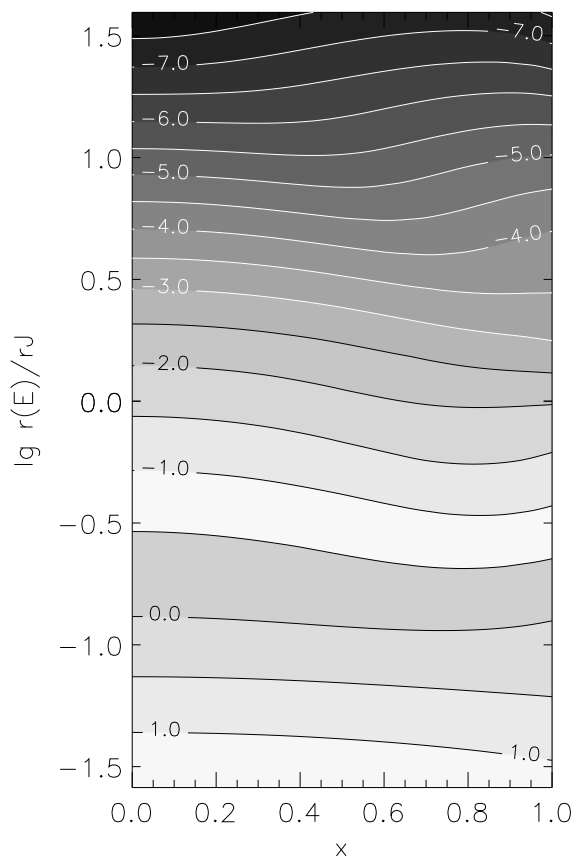}
 \vspace*{-0.4cm}
 \caption[]
    {Distribution function in the energy--circularity plane derived
     for NGC 6703, in the luminous plus dark matter potential with
     $v_c(R_m)=231\kms$ shown as the middle solid line in the upper
     panel of Fig.~\ref{kinematicfits}. Energy is specified by the
     value of $r(E)$, where $E=\Phi(r)$, in units of the fitted Jaffe
     radius $r_J$, and the \df\ is given in
     arbitrary logarithmic units. The last measured kinematic data point
     is located at $\lg r(E)/r_J = 0.23$.}
 \label{exampledf}
\end{figure}

From Fig.~\ref{kinematicfits} we conclude that the stellar
distribution function in NGC 6703 is near-isotropic at the centre and
then changes to slightly radially anisotropic at intermediate radii
($\beta\equal 0.3-0.4$ at 30'', $\beta\equal 0.2-0.4$ at 60''). It is
not well-constrained near the outer edge of the data, where formally
$\beta\equal -0.5 - +0.4$, depending on the correct potential in the
allowed range.  However, the models with large asymptotic halo
circular velocities shown in Fig.~\ref{kinematicfits} appear less
plausible, because they are the models with the most rapidly
increasing velocity dispersions outside $R\simeq 60''$. The same
models also show the most rapid increase in tangential anisotropy at
and beyond $R_m=78''$, which again appears a priori implausible,
because it implies rapid changes in the \df\ just outside the observed
range. The combined signature of both effects is strongly reminiscent
of Fig.~\ref{fits_235m}, where it was clearly an artifact of the
limited radial range of the data.  If this assumption is correct, one
would again conclude that lower--$v_c(R_m)$ models in the formal range
are favoured. Fig.~\ref{exampledf} shows the recovered \df\ for
the potential with $v_c(R_m)=231\kms$ in Fig.~\ref{kinematicfits}.
Note, however, that all models shown in
Fig.~\ref{kinematicfits} except the self--consistent one are formally
consistent with the presently available data.

\begin{figure}
\centering
\getfig{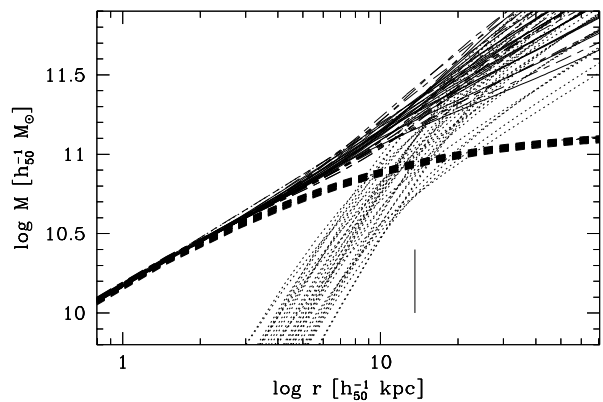}
 \vspace*{-0.4cm}
 \caption[M(r) of dynamical models for NGC 6703]
    {Luminous, dark, and total mass as a function of radius for the range
     of acceptable models of NGC 6703, according to Fig.~\ref{chisqvc}
     (short dashed, dotted, and dash--dashed or full lines, respectively). 
     Mass distributions in which a \df\ with $\chish\le 1.12$ (including $87\%$
     of the distribution in Fig.~\ref{distrchisq}, $1.5\sigma$) are coded by
     full lines, those with $\chish\le 1.28$ (including $95\%$, $2.0 \sigma$)
     by dash--dashed lines. The vertical line denotes the position of
     the last kinematic data point. At this radius, the luminous mass
     fails by at least a factor of $1.6$.}
 \label{mr}
\end{figure}

\begin{figure}
\centering
\getfig{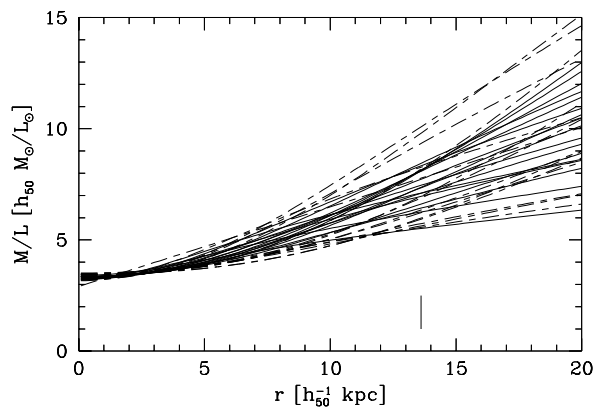}
 \vspace*{-0.4cm}
 \caption[M/L of dynamical models for NGC 6703]
    {The B-band mass--to--light ratio of NGC 6703. The solid and dash--dashed
     lines (coding as in Fig.~\ref{mr}) are
     derived from the dynamical models that span the range of acceptable
     $v_c(R_m)$ in Fig.~\ref{chisqvc}. The central mass--to--light ratio
     is $\Upsilon=3.3$, that at the position of the last kinematic
     data point at 78'' (vertical bar) is in the range of
     $\Upsilon=5.3-10$. }
 \label{allml}
\end{figure}

From the constraints on the circular velocity $v_c(R_m)=210-290\kms$
the range in mass inside $R_m=13.5\,\hfifty^{-1} \kpc$ is
\begin{equation}
M(<R_m) = 1.6-2.6 \times 10^{11} \hfifty^{-1} \msun,
\end{equation}
 where $h_{50}\equiv H_0 / 50 {\rm km/s/Mpc}$. The total mass in stars
inside this radius is $8-9 \times 10^{10} \msun$, assuming constant
mass-to-light ratio $\Upsilon$ and a maximum stellar mass model,
and taking an average value from the
models consistent with the kinematic data. The radial run of the
luminous, dark, and total mass is shown in Fig.~\ref{mr} for the
models that span the allowed range according to Fig.~\ref{chisqvc}.
After dividing by the luminosity $L_B(r)$ for the stars, the
mass-to-light ratios shown in Fig.~\ref{allml} result.  Between the
centre and the last data point $r\equal 78''\ssimeq 2.6R_e$, the
mass-to-light ratio of NGC 6703 rises by a factor of $1.6-3$.

\subsection{Uncertainties}

There are a number of possible sources of systematic error which would
affect the mass--to--light ratio derived for NGC 6703. Most of the
errors so introduced are probably small compared to the considerable
uncertainty arising from the kinematic measurement errors and limited
radial sampling, discussed above. One systematic error on the absolute
mass-to-light ratios in NGC 6703 comes from the uncertainty in the
distance, although this does not change the ratio of outer to central
values. A further systematic effect on this ratio can be introduced by
the sky brightness level: If this is increased by $2-3\%$, the fitted
$r_J$ decreases to 35''. To the extent that the outermost kinematic
data point at 78'' (which then moves to larger $r/r_J$) is in the flat
part of the circular rotation curve, the inferred M/L changes only to
second order because the luminosity inside 78'' remains essentially
unchanged. The same is true if the sky value is decreased by $1\%$, in
which case the fitted $r_J$ increases to $54''$.

In the previous analysis, we have ignored a possible small rotation in
the outer parts of NGC 6703 (perhaps $\sim 20-30\kms$ at $R\gta 50''$,
but the errors are large; cf.\ Fig.~\ref{obsvns}). The simplest
possible estimate of the effect of this rotation on the derived masses
is to replace $\sigma\simeq 105\kms$ in this radial range by
$(\sigma^2+v^2)^{1/2}\simeq 110 \kms$. This gives a factor of $1.1$,
neglecting changes in the model structure that would result because
the central kinematics remain unchanged. 

Next we consider the possibility that NGC 6703 may contain a
face-on extended disk (see de Vaucouleurs, de Vaucouleurs and Corwin
1976). From an $R^{1/4}$-law plus disk decomposition, we estimate that
the contribution of such a disk in the region where we model the
kinematics could be up to 10-20\%.  In this case we expect the
velocity dispersion to be decreased and the $h_4$ coefficient to
become more positive where the disk contributes significant light
(Dehnen \& Gerhard 1994; see NGC 4660 as an example in BSG), 
most likely in the outer parts.

Similarly, it is conceivable that NGC 6703 is in reality slightly
triaxial and is seen from a special direction so as to appear E0. The
likelihood of this is the smaller, the more triaxial the intrinsic
axial ratios are; thus slightly triaxial shapes are the most plausible
ones.  Again this will imply some extra loop orbits seen nearly
face-on, similarly increasing $h_4$ and decreasing $\sigma$.

In both cases, we therefore expect the spherical component in NGC 6703
to have lower $h_4$ and larger $\sigma$ than the measured values. A
similar analysis of such kinematics would, according to the discussion
in Section 3, lead to a model with greater tangential anisotropy at
large radii, with the mass distribution less affected. Recall that
decreasing the mass at large $r$ in a spherical model lowers {\it
both} $\sigma$ {\it and} $h_4$.

\section{Discussion and Conclusions}

This study is part of an observational and
theoretical program aimed at understanding the mass distribution and
orbital structure in elliptical galaxies. In the following we
first discuss general results on potential and anisotropy
determination, and then proceed to the specific case of NGC 6703.

\subsection{Velocity profiles and anisotropy and mass}

The analysis of the \vp s of simple dynamical models in Section 3 has
broadly confirmed the conclusions of G93.  At large radii, where the
luminosity profile falls rapidly, the \vp s are dominated by the stars
at tangent point. Then radially (tangentially) anisotropic \df s can
be recognized by more peaked (more flat-topped) \vp s with more
positive (negative) $h_4$ than for the isotropic case. Increasing
$\beta$ at constant potential thus lowers $\sigma$ and increases
$h_4$.  On the other hand, increasing the mass of the system at large
$r$ at constant anisotropy, increases {\sl both} the projected
dispersion and $h_4$. This suggests that by modelling $\sigma$ and
$h_4$ both mass $M(r)$ and anisotropy $\beta(r)$ can be constrained.

In practical applications, such an analysis is complicated by a number
of factors. Radial orbits at large radii may lead to increased central
velocity dispersions and flat--topped central \vp s (already pointed
out by Dejonghe 1987). The former effect can be compensated by a
decrease in the stellar mass--to--light ratio. The latter is
independent of this, but can be compensated by changes to the
distribution function in the inner parts of the galaxy (as in a number
of cases studied in Sections 4, 5).  A more serious uncertainty is
introduced by the possibility of significant gradients in the orbit
population across the radii of interest. For example, a population of
high energy radial orbits with pericentres in a limited radial range
may mimic tangential anisotropy there. In many cases it will be
possible to exclude such a population of orbits by its effects on the
\vp s at exterior radii, i.e., by simultaneously analysing a number of
observed \vp s.  Yet this is least possible precisely at the largest
observed radii, where mass determination is most interesting. Thus
this chain of argument suggests (correctly, see below) that the
largest uncertainty in determining masses and anisotropies in
ellipticals from \vp--data is the finite radial extent of
these data.

To analyze realistic data we have constructed an algorithm by which
the distribution function and potential of a spherical galaxy can be
constrained directly from its observed $\sigma$ and $h_4$--profiles.
To assess the significance of the results obtained, we have tested the
algorithm on Monte Carlo--generated data sets tuned to the spatial
extent, sampling, and observational errors as measured for NGC
6703. From such data, the present version of the algorithm recovers a
smooth spherical \df, $\sim 70\%$ of the time, to an \rms\ level of
better than $\sim 12\%$ inside three times the radius of the outermost
kinematic data point.

We have used this algorithm to study quantitatively the degree to
which the gravitational potential can be determined from such data.
Our main conclusion is that velocity profile data with presently
achievable error bars already constrain the gravitational
potentials of elliptical galaxies significantly.  In particular,
constant--M/L models are relatively easy to rule out once the data
extend beyond $2R_e$. The examples that we have studied in detail,
tuned to the NGC 6703 data, certainly belong to the less favourable
cases, because in this galaxy the dispersion profile is falling.

A good way to parametrize the results is in terms of the true circular
velocity $v_c(R_m)$ at the radius of the outermost data point,
$R_m$. With presently available data, $v_c(R_m)$ can be determined to
a precision of about $\pm \lta 50\kms$. This will improve when
high--quality data at several $R_e$ become available, of the kind
expected from the new class of 10m telescopes. Apart from the fact
that smaller error bars will decrease the formally allowed range in
$v_c(R_m)$, tests show that this range often includes high--$v_c(R_m)$
models which become rapidly tangentially anisotropic just outside the
data boundary. These (not very plausible) models can be eliminated
with data extending to larger radii.

On the other hand, the detailed form of the true circular velocity
curve is much harder to determine than $v_c(R_m)$. Conspiracies in the
\df\ are possible that minimize the measurable changes in the line
profile parameters. Our tests showed that two potentials differing by
just the value of the halo core radius could not be distinguished even
with very good data out to $6R_e$. Thus some uncertainty will remain
in practise, regardless of whether or not in theory the potential is
uniquely determined from the projected \df\ $N(r,\vpar)$.

A similar picture holds for the related determination of the anisotropy
of the \df. For the present error bars in the data, $\beta(r)$ is
relatively well--determined out to about half the limiting radius
of the observations. Near the edge of the data, uncertainties can
be large depending on the gravitational potential (recall that in
a fixed spherical potential, the \df\ is uniquely determined by
the complete projected \df\ $N(r,\vpar)$). Again the unknown
nature of the orbits beyond the last data point has a large part
in this uncertainty.

Because the largest uncertainties in determining masses and
anisotropies from \vp s occur near the outer radial limit of these
data, the combination of the type of analysis presented here with
other information (e.g., from X-ray data) will be particularly
powerful.

\subsection{The dark halo of NGC 6703}

Fig.~\ref{chichi} shows that {\sl no self-consistent model will fit
the kinematic data for NGC 6703}. Our non--parametric best--estimate
self--consistent model is inconsistent with the data at the
$>99\%$--level (Figs.~\ref{chisqvc}, \ref{distrchisq}). With
self-consistent models, either the velocity dispersion profile is
matched reasonably well, but then the line profiles cannot be
reproduced, or, when the $h_4$-profile is fitted accurately, the
dispersion profile is poorly matched.  The cure for the discrepancy is
to raise {\sl both} $\sigma$ and $h_4$ at large radii.  Thus, as
discussed above, we require extra mass at large $r$.  NGC 6703 must
have a dark halo.

We have next derived constraints on the parameters of this halo as
follows. The luminous component is assigned a constant mass-to-light
ratio $\Upsilon$, chosen maximally such that the stars contribute as
much mass in the center as is consistent with the kinematic data. Our
parametric model for the halo incorporates a constant density core,
and its parameters (core radius $r_c$ and asymptotically constant
circular velocity $v_0$) are chosen such the halo adds mass mainly in
the outer parts of the galaxy if that is necessary.  We call these
models {\sl maximum stellar mass} models (analogous to the maximum
disk hypothesis in the analysis of disk galaxy rotation curves).

We find that maximum stellar mass models in which the luminous
component provides nearly all the mass in the centre fit the data
well. In these models, the total luminous mass inside the limiting
observational radius $R_m\equal 78''\equal 13.5\,\hfifty^{-1} \kpc$ is
$9 \times 10^{10} \hfifty^{-1} \msun$, corresponding to a central
B--band mass--to--light ratio $\Upsilon=3.3 \hfifty
\msun/\lsun$. According to Worthey's (1994) models, this is a rather
low value for the stellar population of an elliptical galaxy and would
point to a relatively low age (5 Gyrs) and/or low metallicity (less
than solar). However, the galaxy has a color $(B-V)_0=0.93$ and a
central line index $Mg_2=0.280$ (Faber et al.\ 1989) which are typical
for ellipticals of similar velocity dispersion.

A larger value of $H_0$ could increase the M/L value and alleviate the
demands on the stellar populations. However, the distance used here
(36 Mpc) includes a correction for the large inferred peculiar
velocity of the galaxy. If we had used a distance based on the larger
radial velocity in the CMB frame, our derived M/L would be even lower.
It is also implausible that the low central value of M/L stems from
the contribution of a young stellar population in a disk component,
which we estimate cannot be larger than $20\%$ of the total light (see
above). Thus we conclude that the dark halo in NGC 6703 is unlikely to
have higher central densities than inferred from our maximum stellar
mass models, because otherwise the M/L of the stellar component would
be reduced to implausibly small values.

In a recent preprint, Rix \etal (1997) have analyzed the velocity
profiles of the E0 galaxy NGC 2434 with a linear orbit superposition
method. This galaxy provides an interesting contrast to NGC 6703
because it has an essentially flat dispersion profile. Its kinematics
are likewise inconsistent with a constant--M/L potential, but are
well--fit by a model with $v_c\equal {\rm const}$. This can be
interpreted as a maximum stellar mass model in the sense defined
above, in which the luminous component with maximal $\Upsilon$
contributes most of the mass inside $R_e$. The kinematics of NGC 2434
are also well--fit by a range of specific, cosmologically motivated
mass models which, if applicable, would imply lower $\Upsilon$ and
significant dark mass inside $R_e$. In NGC 6703, a model with
$v_c\equal {\rm const}$ is formally consistent with the present data
(within $2\sigma$), but it is not a very plausible fit at large $R$
and requires large anisotropy gradients between 40'' and 70''. It will
be interesting to see whether future studies confirm differences
between the shapes of the true circular velocity curves of elliptical
galaxies.

Because of the falling dispersion curve in NGC 6703, we can determine
only one of the halo parameters (The halo parameters of the most
probable potentials lie in a band extending from low $v_0$ and low
$r_c$ to high $v_0$ and high $r_c$.)  However, the circular velocity
$v_c(R_m)$ at the data boundary is relatively well--determined for all
these models.  Thus we find (Fig.~\ref{chisqvc}) that the true
circular velocity of NGC 6703 at 78'' is $v_c(R_m)=250\pm 40\kms$
(formal $95\%$ confidence interval).  Tests on pseudo data have shown
that this range often includes high--$v_c(R_m)$ models which become
rapidly tangentially anisotropic just outside the data boundary. Such
models may not be very plausible, so the lower values in the quoted
range of $v_c(R_m)=250\pm 40\kms$ may be the more probable ones.

Thus, at $R_m\equal 78''\equal 13.5\,\hfifty^{-1} \kpc$ the total mass
enclosed is $M(<R_m) = 1.6-2.6 \times 10^{11} \hfifty^{-1} \msun$, and
the integrated mass--to--light ratio out to this radius is
$\Upsilon=5.3-10$, corresponding to a rise from the center by at least
a factor of $1.6$. We have already noted that NGC 6703 is an
unfavourable case because of its falling dispersion curve. The fact
that relatively small variations in $\Upsilon$ can nonetheless be
detected shows the power of the method. Note that a scheme
based on the analysis of the line of sight velocity dispersions alone
(Binney, Davies, Illingworth 1990, van der Marel 1991) would conclude
that constant mass--to--light ratio models can provide good fits.

The stellar distribution function in NGC 6703 is near-isotropic at the
centre and then changes to slightly radially anisotropic at
intermediate radii ($\beta\equal 0.3-0.4$ at 30'', $\beta\equal
0.2-0.4$ at 60''). It is not well-constrained near the outer edge of
the data, where formally $\beta\equal -0.5 - +0.4$, depending on the
correct potential in the allowed range. Models near the lower
end of this range may be consistent with the data only because of the
limited radial extent of the measurements.
\subsection{Conclusions}

In summary, we have shown that the mass distribution $M(r)$ and
anisotropy structure $\beta(r)$ for spherical galaxies can {\it both\/
} be constrained from \vp\ and velocity dispersion measurements. NGC
6703 must have a dark halo, contributing about equal mass at $2.6 R_e$
as do the stars. The circular velocity at the last kinematic data
point (78'') is constrained to lie in the range $250\pm 40\kms$ at
$95\%$ confidence. The anisotropy of the stellar orbits changes from
near-isotropic at the center to slightly radially anisotropic at
intermediate radii, and may be either radially or tangentially
anisotropic at 78''. With more extended and more accurate data it will
be possible to considerably narrow down these uncertainties.

If the results for this galaxy are typical, they suggest that also in
elliptical galaxies the stellar mass dominates at small radii, and the
dark matter begins to dominate at radii around $10\kpc$. It is
important to obtain extended kinematic data and do a similar analysis
for a number of elliptical galaxies.  When we know the systematics and
the spread in the circular velocity curves and anisotropy profiles for
a sample of ellipticals, we will have an important new means for
testing the currently popular formation theories.

%
%
\section*{Acknowledgments} 

We thank U.~Hopp for providing us with a CCD frame of NGC 6703, and
David Merritt for helpful discussions on regularization methods.
We thank the referee for his rapid and constructive comments,
especially on the revised version, and the editorial staff for
their relentless efforts to secure his reports. We
acknowledge financial support by the Deutsche Forschungsgemeinschaft
under SFB 328 and SFB 375 and by the Schweizerischer Nationalfonds
under grants 21-40464.94 and 20-43218.95. OEG also acknowledges
a Heisenberg fellowship while at Heidelberg.

\section*{Appendix A: Library of anisotropic spherical distribution functions}

To understand the connection between anisotropy structure and
observable line profile shapes we have constructed a number of
spherical distribution functions of the quasi--separable form
(Gerhard 1991, G91)
\leqnam{\eqdeff}
\begin{equation}
f(E,L) = g(E) \mathsp h(x),
\end{equation}
where the variable $x$ depends on both energy and angular momentum:
\leqnam{\eqdefx}
\begin{equation}
x = {L \over L_0 + L_c(E)}.	
\end{equation}
 $L_0$ is an angular momentum constant, or equivalently, an anisotropy
radius times a characteristic velocity; $L_c(E)$ is the angular
momentum of the circular orbit at energy $E$. \df s of the form
\eqdeff) have the following properties: (i) the circularity
function $h(x)$ has the effect of shifting stars between orbits of
different angular momenta on surfaces of constant energy, while
$g(E)$ controls the distribution of stars between energy
surfaces. (ii) For the most bound stars $L < L_c(E) \ll L_0$; thus
the model becomes isotropic in the centre unless $L_0\equal 0$. (iii)
For loosely bound stars $ L \sim L_c(E) \gg L_0$, i.e., the angular
momentum distribution becomes a function of circularity $L/L_c(E)$
which is one-to-one related to orbital eccentricity. Outside the
anisotropy radius, the \df\ \eqdeff) therefore corresponds to an
energy-independent orbit distribution with constant anisotropy, radial
or tangential. 

In these models $h(x)$ is an assigned function built into
the model to achieve the desired anisotropy (orbit distribution).
Radially biased distribution functions correspond to circularity functions
$h(x)$ decreasing with $x$; for example
\leqnam{\halfa}
\begin{equation}
 h(x) = h_\alpha(x) \equiv \left( 1 - x^2 \right)^\alpha.
\end{equation}
The family can also be used to construct tangentially
anisotropic models, such as
\leqnam{\htang}
\begin{equation}
 h_{c,\alpha}(x) = c + (1-c) \left[ 1 - (1-x^2)^\alpha \right].
\end{equation}
 In these tangentially anisotropic models one cannot choose $h(0)=0$
unless $L_0=0$, otherwise the density at $r\equal 0$ would be zero.
Of course, other forms for $h(x)$ are possible, such as Gaussians.

\begin{figure}
\centering
\getfig{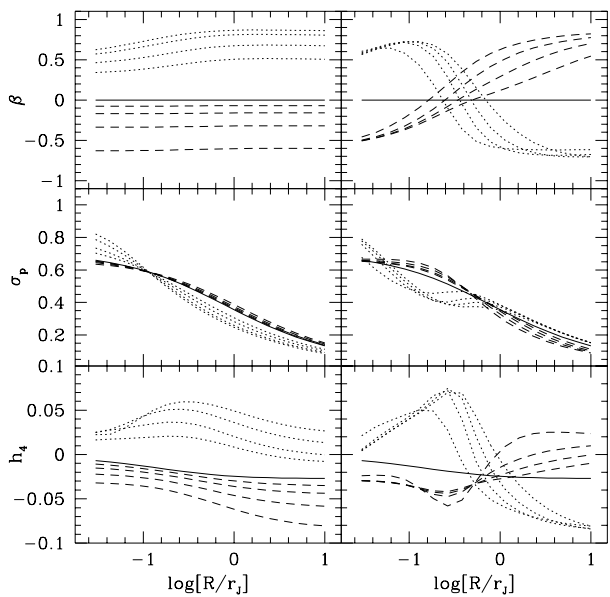}
 \vspace*{-0.4cm}
 \caption[animodels]
    {Anisotropy parameter $\beta(r)$, projected velocity dispersion
     $\sigma_p$, and line profile shape parameter $h_4$ for several
     families of anisotropic \df s. Left: scale--free radially
     anisotropic (dotted) and tangentially anisotropic (dashed lines)
     in the potential of a self--consistent Jaffe sphere. For these
     models, circularity functions of the type \halfa) and \htang)
     were used with different $\alpha$ and $c$, respectively.
     The isotropic model is shown for reference
     (solid line). Right: Families of models with anisotropy changing
     from radial to tangential (dotted) or from tangential to radial
     (dashed lines). These were constructed from the same circularity
     functions and a weight factor as in eq.~(23).}
 \label{figatwo}
\end{figure}

Given the assigned function $h(x)$, the integral equation for
$\rho(r)$ in terms of $f(E, L^2)$ is solved for the derived
function $g(E)$; see G91 and Jeske (1995). Fig.~\ref{figbasis}
shows line profile shape parameters for representative \df s
constructed in this way. Fig.~\ref{figbasis} shows the anisotropy
profiles of two sets of tangentially anisotropic models. Here the
density of stars has been taken to be that of a Jaffe sphere, and the
potential in which the stars orbit is either the self--consistent
potential or one of the mixed stars plus halo potentials used in
Sections~4,~5. Sequences like that in Fig.~\ref{figbasis} are used as
basis functions in the non--parametric analysis in Section~5.

Models whose anisotropy changes from radial to tangential or vice
versa were constructed by linearly combining the above circularity
functions with energy--dependent coefficients. In this way one
obtains \df s of the more general form
\leqnam{\eqdefgen}
\begin{equation}
f(E,L) = g(E) \mathsp h(E,x).
\end{equation}
For self--consistent Jaffe models we used energy--dependent coefficients
$\mu(E)$ of the following form:
\leqnam{\defmueps}
\begin{equation}
\mu(E) = {1\over 2} + {1\over \pi} \arctan{\left[{k\over 2}
	{E^2-\=E^2\over E^2}\right] }.
\end{equation}
The parameters $\=E$ and $k$ determine the orbital energy
near which the anisotropy transition occurs, and the width of the
transition. A similar function $\mu(E)$ was used for models
in halo potentials. Figure \ref{figatwo} shows the intrinsic and projected
properties of a number of \df s of this kind, constructed in the
self--consistent potential of a Jaffe sphere. Notice the wide
variety of kinematical profiles that can be constructed in this
way.

\section*{Appendix B: Transforming to linear kinematic data}

In velocity line profile measurements, the depth of an absorption line
in a spectral resolution element is assumed to be proportional to the
number of stars with line-of-sight velocities corresponding to this
wavelength interval.  The line-of-sight velocity distribution measured
from the line profiles is a discretized function linearly related to
the underlying \df\ [cf.~eq.~\projdf)]. This linearity is lost when
line profile measurements are represented by the quantities $v$,
$\sigma$, $h_3$, $h_4$ --- these quantities are obtained by a
least-square {\sl fit} of a Gauss-Hermite series to the observed line
profile.

To re-express the observed kinematics in terms of quantities that
depend linearly on $f$ we proceed as follows.  Consistent with the
assumption of spherical symmetry, we set the mean streaming velocity
$\= v$ and all odd velocity profile moments to zero.  Next, we obtain
an estimate for the velocity dispersion (second moment) $\=\sigma^2$,
by integrating over the line profile specified by $(\sigma, h_4)$; for
negative $h_4$, until it first becomes zero. For small $h_4$, the
linear correction formula $\=\sigma\equal \sigma (1+\sqrt{6}h_4)$
holds (van der Marel \& Franx 1993), this correction results in
$\=\sigma\!>\!\sigma$ for peaked profiles with $h_4\!>\!0$. The
numerical correction from integrating over the velocity profile also
has this property (BSG).

From the measured $h_4(R_i)$, we compute new even Gauss-Hermite moments
$s_n(R_i;\s\sigma)$ by expanding the series
\begin{equation}
  {\cal L}(R_i,\vpar) = {\cal L}_0 \sum_{j=0,4}
	h_j(R_i)\, H_j(x)\, \exp(-x^2/2)
\end{equation}
 ($h_0\equal1$, $h_1\equal h_2\equal h_3\equal 0$) with
$x\equiv\vpar/\sigma(R_i)$ as
\leqnam{\newsn}
\begin{equation}
  {\cal L}(R_i,\vpar) = \!\sum_{n=0,2,4,...} \!\!\! N_n\, s_n 
	\, H_n(\s x)\, \exp(-\s x^2/2),
\end{equation}
 where $\s x\equiv\vpar/\s\sigma(R_i)$. Here $H_n$ are Hermite
polynomials, the $N_n$ are normalization constants (G93), and
$\s\sigma(R_i)$ are fiducial scaling velocities generally different
from $\=\sigma(R_i)$. In practice, we have found it convenient to take
for $\s\sigma(R_i)$ the velocity dispersions $\sigma_{\rm iso}(R_i)$
of the isotropic model in the given potential $\Phi(r)$ with the same
stellar density as the galaxy being analyzed. The $s_n(R_i;\s\sigma)$
are estimates for the Gauss-Hermite {\sl moments} related to the
true velocity profile by
\leqnam{\sniso}
\begin{eqnarray}
  s_n(R_i;\s\sigma)
	&=& \left(2^{n-1} n!\right)^{-1/2} \\
	   && \times \int_{-\infty}^\infty \! d\vpar \, H_n(\vpar/\s\sigma)
	\, \exp(-\vpar^2/2\s\sigma^2) \, {\cal L}(R_i,\vpar). \nonumber 
\end{eqnarray}

For a theoretical model, we obtain corresponding moments by inserting
the projected \df\ $N(R_i,\vpar)$ from \projdf) into eq.~\sniso)
instead of ${\cal L}(R_i,\vpar)$, using the same
$\s\sigma(R_i)$. Clearly, the new $s_n$-moments of the composite model
are linear in the $a_k$, i.e., 
\begin{equation}
   s_n(R_i;\s\sigma) = \sum_{k=1,K} a_k s_n^{(k)}(R_i;\s\sigma)
\end{equation}
 with the $s_n^{(k)}$ corresponding to the respective $f_k$.


\begin{thebibliography}{}

\bibitem[Awaki \etal, 1994]{Awaki94} Awaki H., \etal, 1994, \pasj{46}{L65}
\bibitem[Arnaboldi \etal, 1994]{a94} Arnaboldi M., Freeman K.C., Hui X.,
  Capaccioli M., Ford H., 1994, Messenger {\bf 76}, 40
\bibitem[Bender, 1990]{ben90} Bender R., 1990, \aa{229}{441}
\bibitem[Bender, Saglia \& Gerhard, 1994]{bsg94} Bender R., Saglia, R.,
	 Gerhard, O.E., 1994, \mn{269}{785} (BSG)
\bibitem[Binney, 1978]{bin78} Binney J.J., 1978, \mn{183}{501}
\bibitem[Binney, Davies \& Illingworth 1990]{bdl90}Binney, J.J., Davies,
	R.L., Illingworth, G.D., 1990, \apj{361}{78}
\bibitem[Binney \& Mamon, 1982]{bm82} Binney J.J., Mamon G.A., 1982, 
  	\mn{200}{361}
\bibitem[Carollo \etal, 1995]{c95} Carollo C.M., de Zeeuw P.T., van der
  	Marel R.P., Danziger I.J., Qian E.E., 1995, \apjl{441}{L25}
\bibitem[Dehnen \& Gerhard 1994]{dg94} Dehnen W., Gerhard O.E., 1994, 
  \mn{268}{1019}
\bibitem[Dejonghe, 1987]{dej87} Dejonghe H., 1987, \mn{224}{13} 
\bibitem[Dejonghe \etal, 1996]{dej96} Dejonghe H., de Bruyne V.,
	Vauterin P., Zeilinger W.W., 1996, \aa{306}{363}
\bibitem[Dejonghe \& Merritt, 1992]{dm92} Dejonghe H., Merritt D., 1992, 
  \apj{391}{531}			
\bibitem[de Vaucouleurs \etal, 1976]{dvplus76} de Vaucouleurs G., 
  de Vaucouleurs A., Corwin H.G., 1976, Second Reference 
  Catalogue of Bright Galaxies, Univ. Texas Press, Austin
\bibitem[Faber \etal, 1989]{fab89} Faber S.M., Wegner G., Burstein D., 
  Davies R.L., Dressler A., Lynden-Bell D., Terlevich R.J., 1989, 
  \apjs{69}{763}
\bibitem[Franx, van Gorkom \& de Zeeuw, 1994]{fgz94} Franx M., van
  Gorkom J.H., de Zeeuw T., 1994, \apj{436}{642}
\bibitem[Fricke, 1952]{fri52} Fricke W., 1952, Astr.~Nachr.\ 280, 193 
\bibitem[Gerhard, 1991]{ger91} Gerhard O.E., 1991, \mn{250}{812} (G91)
\bibitem[Gerhard, 1993]{ger93} Gerhard O.E., 1993, \mn{265}{213} (G93)
\bibitem[Grillmair \etal, 1994]{g94} Grillmair C.J., Freeman K.C., Bicknell
  G.V., Carter D., Couch W.J., Sommer-Larsen J., Taylor K., 1994,
  \apj{422}{L9}
\bibitem[Hanson \& Haskell 1981]{Hanson, Haskell, 1981}
  Hanson R.J., Haskell K.H., 1981, Math.\ Programm.\  21, 98.
\bibitem[Jaffe, 1983]{jaf83} Jaffe W., 1983, \mn{202}{995}
\bibitem[Jeske, 1995]{jes95} Jeske, G., 1995, PhD Thesis, University
	of Heidelberg
\bibitem[Jeske \etal, 1996]{jes96} Jeske, G., Gerhard, O.E., Bender,
  R., Saglia, R.P., 1996, Proc. IAU 171, eds. Bender, R., Davies, R.L., p. 397
\bibitem[Kim \& Fabbiano, 1995]{kf95} Kim D.-W., Fabbiano G., 1995, 
  \apj{441}{182}
\bibitem[Kochanek \& Keeton, 1997]{kk97} Kochanek C.S., Keeton C.R.,
  1997, in The Nature of Elliptical Galaxies, $2^{nd}$ Stromlo Symposium,
  eds.\ Arnaboldi M., Da Costa G.S., Saha P., 1997, ASP 116, 21.
\bibitem[Maoz \& Rix, 1993]{mr93} Maoz D., Rix H.-W., 1993, \apj{416}{435}
\bibitem[Merritt, 1985]{m85} Merritt D., 1985, \aj{90}{1027}
\bibitem[Merritt, 1993]{m93} Merritt D., 1993, \apj{413}{79}
\bibitem[Osipkov, 1979]{o79} Osipkov L.P., 1979, Pis'ma Astr.\ Zh.\ {\bf 5}, 77
\bibitem[Press \etal, 1986] {press} Press, W.H., Flannery, B.P., Teukolsky,
  S.A., Vetterling, W.T., Numerical Recipes (Cambridge: Cambridge
  University Press)
\bibitem[Rix \etal 1997]{rixetal97} Rix H.-W., de Zeeuw P.T., Carollo
  C.M., Cretton N., van der Marel R.P., 1997, preprint
\bibitem[Saglia \etal, 1993]{s93} Saglia R.P.\ \etal, 1993, \apj{403}{567}
\bibitem[Saglia \etal, 1997b]{s97a} Saglia, R.P., Bender, R., Gerhard,
  O.E., Jeske, G., 1997a, in Dark and Visible Matter
  in Galaxies and Cosmological Implications, eds.\ Persic M., Salucci P.,
  ASP 117, 113.
\bibitem[Saglia \etal, 1997b]{s97b} Saglia R.P.\ \etal, 1997b, ApJS
 	{\bf 109}, 79
\bibitem[van der Marel, 1991]{vdm91} van der Marel, R., 1991, \mn{253}{710}
\bibitem[van der Marel \& Franx, 1993]{vmf93} van der Marel R.P.,
	 Franx M., 1993, \apj{407}{525}
\bibitem[Worthey 1994]{w94}Worthey, G. 1994, ApJS, {\bf 95}, 107
\end{thebibliography}
\end{document}